\documentclass[%
 superscriptaddress,
 twocolumn,
 nofootinbib,
 amsmath,amssymb,
 aps,
 prd,
]{revtex4-2}

\usepackage[colorlinks]{hyperref}
\usepackage{tabularx}
\usepackage{graphicx}
\usepackage{amssymb,amsmath,bm,tensor,braket}
\usepackage{xcolor}
\usepackage[varg]{txfonts}
\usepackage{enumerate}
\usepackage{mathtools}
\usepackage[capitalize]{cleveref}
\usepackage[utf8]{inputenc}
\usepackage[normalem]{ulem}
\usepackage{dcolumn}

\usepackage{subfigure}
\usepackage[normalem]{ulem}

\newcommand*\DL{\ensuremath{D_{\rm L}}}
\newcommand*\calM{\ensuremath{\mathcal{M}}}
\newcommand*\ii{\ensuremath{\mathop{}\!\mathrm{i}}}
\newcommand*\rmd{\ensuremath{\mathop{}\!\mathrm{d}}}
\newcommand*\SUN{\ensuremath{{\rm M}_{\odot}}}

\begin{document}

\title{Including higher harmonics in gravitational-wave parameter estimation and cosmological implications for LISA}

\newcommand{\IFAA}{\affiliation{Institute for Frontiers in Astronomy and Astrophysics, Beijing Normal University, Beijing 102206, China}}
\newcommand{\BNU}{\affiliation{Department of Astronomy, Beijing Normal
University, Beijing 100875, China}}
\newcommand{\WHU}{\affiliation{School of Physics and Technology, Wuhan University, Wuhan, Hubei 430072, China}}
\newcommand{\ICRR}{\affiliation{Institute for Cosmic Ray Research (ICRR), KAGRA Observatory, The University of Tokyo,
Kashiwa 277-8582, Japan}}
\newcommand{\HIAS}{\affiliation{School of Fundamental Physics and Mathematical Sciences, Hangzhou Institute for Advanced Study, UCAS, Hangzhou 310024, China}}

\author{Yi Gong}\BNU\WHU\ICRR

\author{Zhoujian Cao}\IFAA\BNU\HIAS

\author{Junjie Zhao\footnote{corresponding author}} \email[Junjie Zhao: ]{junjiezhao@bnu.edu.cn}\IFAA\BNU

\author{Lijing Shao}
\affiliation{Kavli Institute for Astronomy and Astrophysics, Peking University,
Beijing 100871, China}
\affiliation{National Astronomical Observatories, Chinese Academy of Sciences,
Beijing 100012, China}

\date{\today}

\begin{abstract}
Massive black holes (MBHs) are crucial in shaping their host galaxies. How the
MBH co-evolves with its host galaxy is a pressing problem in astrophysics and
cosmology. The valuable information carried by the binary MBH is encoded in the
gravitational waves (GWs), which will be detectable by the space-borne GW
detector LISA. In the GW data analysis, usually, only the dominant $(2,2)$ mode
of the GW signal is considered in the parameter estimation for LISA.  However,
including the higher harmonics in parameter estimation can break the degeneracy
between the parameters, especially for the inclination angle and luminosity
distance. This may enable the identification of GW signals without
electromagnetic counterparts, known as ``dark sirens''. Thus, incorporating
higher harmonics will be beneficial to resolve the Hubble tension and constrain
the cosmological model. In this paper, we investigate the role of higher
harmonics in the parameter estimation for GWs emitted by binary MBHs. We
demonstrate that including $(3,3)$ mode can lead to a $10^3$-times improvement
in angular resolution and a $10^4$-times improvement in luminosity distance.
Meanwhile, our results indicate that considering higher harmonics increases the
probability of identifying over 70\% host galaxies from $10^{-2}\,\rm{Gpc}^3$
cosmological volume threshold (corresponding $10^5$ host galaxies), while the
probability less than 8\% for only the $(2,2)$ mode. Thus, our results
underscore the importance of including higher modes in the GW signal from binary
MBHs, for LISA at least $(3,3)$ mode.
\end{abstract}

\maketitle


\section{Introduction}
\label{sec:intro}

Properties of massive ($\sim 10^4-10^8\,\SUN$) black holes (MBHs) are crucial to
understanding the evolution of their
hosts~\cite{Volonteri:2010wz,Kormendy:2013dxa,Alexander:2017rvg} and are
relevant to open problems in astrophysics and cosmology, such as dark matter,
vacuum energy, and the early universe~\cite{Barack:2018yly}. Gravitational waves
(GWs) provide a valuable tool for constraining the Hubble constant $(H_0)$,
referred to as ``sirens'', by inferring distance directly from the signals, and
the redshift $z$ is from the host galaxy or the electromagnetic (EM)
counterpart~\cite{Schutz:1986gp,Holz:2005df}.

The Laser Interferometer Space Antenna (LISA), a space-borne GW detector
building on the success of LISA Pathfinder and other ground-based detectors,
will target these GW sources from tens of micro-hertz up to
deci-hertz~\cite{LISA:2022kgy,Bayle:2022hvs,LISA:2017pwj}. By observing these
sources, LISA will provide insights into the evolution of MBHs
from the early universe through to the peak of the star formation era, offering
key information for future studies in astrophysics and cosmology.

LISA sensitives to a different frequency band compared to existing ground-based
detectors, the Laser Interferometer GW Observatory
(LIGO)~\cite{LIGOScientific:2014pky}, Virgo~\cite{VIRGO:2014yos}, and
KAGRA~\cite{KAGRA:2020agh}. Signals from MBHs will last several months to years
in the lifetime of LISA, which will lead to signal modulation effects as LISA
changes its orientation and position during
observations~\cite{Cutler:1997ta,LISA:2017pwj}.  The signal-to-noise ratio
(SNR), a measure of the strength of the signal relative to the noise, of the
signal detected by ground-based detectors is typically $\mathcal{O}(10)$ at
present~\cite{LIGOScientific:2018mvr,LIGOScientific:2020ibl,LIGOScientific:2021usb,LIGOScientific:2021djp}.
However, signals from MBHs detected by LISA will have much higher SNRs of
$\mathcal{O}(10^3)$~\cite{Cutler:1997ta,Berti:2004bd}, almost 100-fold louder.
It makes LISA an invaluable tool for studying these elusive sources of GWs and
opens up new opportunities for insights into their properties and evolution.

The signal modulation effect caused by LISA motion will break the parameter
degeneration, resulting in a high-precision source parameter extraction.
Incorporating higher harmonics into GW data analysis will further break the
parameter degeneration, which may lead to a higher precision parameter
extraction. For GWs emitted by compact binary coalescences (CBCs), the waveform
can be expressed as a linear superposition of spin-weighted spherical harmonic
functions and the waveform component of each
harmonic~\cite{Thorne:1980ru,Blanchet:2013haa}. The dominant mode of the GW
signal is the $(\ell,|m|) = (2,2)$ mode. Subdominant harmonics
such as $(\ell,|m|) = (3,3)$, $(4,4)$, or $(2,1)$ can usually have amplitudes up
to $\sim 10\%$ of the dominant mode~\cite{London:2017bcn}. However, the
subdominant harmonics are usually ignored in the data analysis for LISA.

Incorporating higher harmonics into GW data analysis has become increasingly
important in recent years. There are several binary black hole (BBH) events
detected by GW observatories that show evidence of higher harmonics. Notably,
GW190412~\cite{LIGOScientific:2020stg} is the first event with significant
evidence for higher harmonics, featuring a mass ratio of approximately 3.57 and
a total SNR of about 19. The $(\ell, |m|) = (3,3)$ harmonic mode contributes
about 3 to the SNR. Another event, GW190814~\cite{LIGOScientific:2020zkf}, also
shows evidence of the $(\ell, |m|) = (3,3)$ harmonic mode, with a mass ratio of
approximately 8.93 and a total SNR of about 25. This event is particularly
asymmetric in masses. These two events, along with GW170729 (slightly weaker
evidence)~\cite{Chatziioannou:2019dsz,LIGOScientific:2018mvr}, point towards the
potential for higher harmonics to improve parameter estimation. This means that
these higher harmonics can be detectable in GW signals with large SNRs, and
their inclusion in waveform models can improve parameter estimation. Because of
the high SNR of LISA data, higher harmonics are indispensable in LISA data
analysis.

Higher harmonics are essential for accurately estimating orbit
precession~\cite{Krishnendu:2021cyi}, and can also help to constrain alternative
gravity theories~\cite{Wang:2022apn,Gao:2022hsn}. Ignoring higher harmonics can
lead to significant systematic errors in parameter
estimation~\cite{Varma:2016dnf,Varma:2018mmi}. Several recent studies have
investigated the effects of including higher harmonics on LISA parameter
estimation. When considering the inspiral phase, using higher post-Newtonian
(PN) orders that include higher harmonics can improve LISA angular resolution by
a factor of about $10^2$~\cite{Porter:2008kn,Trias:2007fp,Arun:2007hu}. Higher
harmonics of the ringdown phase may be detected by LISA through
analysis of the ringdown waveform, and subdominant harmonics play a critical
role in source localization~\cite{Baibhav:2020tma}. When analyzing full
waveforms, incorporating higher harmonics can improve the precision of luminosity
distance estimation by a factor of around 50~\cite{Wang:2022apn}, and sky
angular resolution by a factor of approximately $10^3$ in certain specific
systems~\cite{Marsat:2020rtl,Pratten:2022kug}. Many other recent studies also
support the conclusion that including higher harmonics can improve parameter
estimation~\cite{Katz:2021uax,Ng:2022vbz,Iacovelli:2022mbg}.

The precision of localization and distance inference is a crucial factor to
achieve a high-precision $H_0$ measurement.  However, EM counterparts to GW
events are much rarer than events without counterparts, which would lead to poor
localization. To date, only one GW event, GW170817, has provided a $H_0$
measurement using an EM
counterpart~\cite{LIGOScientific:2017vwq,Coulter:2017wya,LIGOScientific:2017zic,LIGOScientific:2017ync,LIGOScientific:2017adf}.
Due to the degeneracy between the luminosity distance and the inclination angle,
the measurement of $H_0$ from GW170817 is not good enough. The GW measurement
 is broadly consistent with the results from Plank~\cite{Planck:2018vyg}
and SH0ES~\cite{Riess:2021jrx} Collaborations, while these two results are
inconsistent at the $\geq 3\,\sigma$ level. GWs without EM counterparts,
referred to as ``dark sirens'', can also be used to constrain $H_0$ in a
statistical
way~\cite{DES:2019ccw,DES:2020nay,LIGOScientific:2019zcs,LIGOScientific:2021aug,Finke:2021aom},
but the constraint on $H_0$ is limited by the uncertainty in the source
location. Incorporating higher harmonics into data analysis is promising to
resolve this conundrum.  The considerable improvement in localization and
distance inference provided by GWs makes them a more powerful tool for
investigating cosmology by locating the host galaxy, and a promising avenue for
resolving the existing $H_0$ tension
problem~\cite{Planck:2018vyg,Riess:2021jrx}. 

Recently, \citet{Yang:2022tig} found GWs emitted by the eccentric compact
binaries in the deci-hertz band can significantly improve the angular
resolution, potentially leading to new opportunities for probing cosmology.
However, the question of the probability of achieving such a dramatic
improvement in the source location -- a crucial factor for studying $H_0$ and
cosmology -- remains unanswered. 

In this paper, we investigate how the parameters such as binary masses, source
location, and higher harmonics impact LISA parameter estimation.  When
considering a face-on or face-off MBH binary with the total mass of $\sim
10^6\,\SUN$, we find the varying mass ratio only light impacts the parameter
extraction. Meanwhile, including higher harmonics will significantly enhance the
precision of luminosity distance and angular resolution by a factor of $10^4$
and $10^3$, respectively.  In the meantime, higher harmonics will improve the
ability to localize the host galaxies of MBH binaries, compared to only
involving $(2,2)$ mode. For instance, if the threshold volume is
$10^{-2}\,\rm{Gpc}$ (excepted number of host galaxies within this volume is
$10^5$), including $(3,3)$ mode will increase this probability from less than
8\% to 70\%.  Thus, Our results suggest including higher modes for LISA data
analysis, at least $(3,3)$ mode.

The rest parts of this paper are organized as follows. In
Sec.~\ref{sec:Gravitational waves in LISA}, we review the GW waveform including
higher harmonics. Meanwhile, we also review the Fisher matrix method to estimate
the measurement precision. In Sec.~\ref{sec:Result}, we exhibit the effect
caused by higher harmonics. In Sec.~\ref{sec:Conclusion}, we make a summary of
our work and provide some concluding remarks. The geometrized units $G=c=1$ are
used throughout this paper.

\section{Gravitational waves detected by LISA}
\label{sec:Gravitational waves in LISA}

The frequency-domain GW strain $\tilde h(f)$ measured by LISA
is~\cite{Cutler:1997ta,Liu:2020nwz},
\begin{eqnarray}
    \label{eqn:hf}
\tilde h(f) = \frac{\sqrt{3}}{2}\left(F^+(f)\tilde h_+(f)+ F^\times(f) \tilde
h_\times(f)\right)e^{-\ii\phi_D(f)}\,, 
\end{eqnarray}
where $\tilde h_+(f)$ and $\tilde h_\times(f)$ are two orthogonal polarizations
of GWs, `plus' and `cross' modes in the transverse-traceless gauge; $F^+(f)$ and
$F^\times(f)$ are the pattern functions of the detector and there is a factor
$\frac{\sqrt{3}}{2}$ because of the equilateral triangle shape.  $\phi_D(f)$ is
the Doppler effect phase caused by LISA motion [see Eq.~\eqref{eqn:phi_D}].

In this paper, we employ the \textsc{IMRPhenomHM}~\cite{London:2017bcn} waveform
template to generate $\tilde h_+(f),\,\tilde h_\times(f)$. \textsc{IMRPhenomHM}
is an inspiral-merger-ringdown waveform with the higher harmonics $(\ell, |m|) =
(3,3),\,(4,4),\,(3,2),\,(2,1),\,(4,3)$. For simplicity, we only consider the
non-spinning BBHs with 7 physical parameters to generate the waveform template.
They are the source-frame component masses $m_1$ and $m_2$, the inclination
angle $\iota$, the reference orbit phase $\phi_0$, the time to coalescence
$t_c$, the luminosity distance $\DL$ and the corresponding cosmological redshift
$z$. The binaries' chirp mass $\calM$ and symmetric mass ratio $\eta$ are
defined as,
\begin{subequations}
    \label{eqn:calM_eta}
        \begin{eqnarray}
            &&\eta =  m_1m_2/M^2 \,,\\
            &&\calM = \eta^{3/5}M \,.
        \end{eqnarray}    
\end{subequations}
Here $M = m_1+m_2$ is the total mass of the binary, and the mass ratio is $q
\equiv m_1/m_2\ge 1$.

Moreover, we use the $\rm{\Lambda CDM}$ cosmology model to obtain the luminosity
distance $\DL$ from the redshift $z$,
\begin{equation}
    \DL(z)=\frac{1+z}{H_0} \int_0^z \frac{\mathrm{d} z^{\prime}}{\sqrt{\Omega_{\Lambda} + \Omega_{\rm m}\left(1+z^{\prime}\right)^3}} \,.
\end{equation}
Here, we take this cosmology model with the Hubble constant $H_0 = 67.4 \,
\rm{km\ s^{-1} Mpc^{-1}}$, the matter density parameter $\Omega_{\rm m} = 0.315$
and the dark energy density parameter $\Omega_\Lambda =
0.685$~\cite{Planck:2018vyg}.

In this paper, we fix all the BBH sources located at $z=1$ for illustration.

\subsection{LISA pattern functions}
\label{subsec:pattern functions}

The pattern functions depend on the sky location in the detector frame
$\left(\hat\theta_S,\hat\phi_S\right)$ and the polarization angle of GWs
$\left(\hat{\psi}\right)$ of the GW source~\cite{Cutler:1997ta},
\begin{subequations}
    \label{eqn:PF}
    \begin{eqnarray}
        \label{eqn:F+A}
        F^+_A\left(\hat\theta_S,\hat\phi_S,\hat{\psi}\right) && =   \frac{1}{2}
        \left(1+\cos^2\hat\theta_S\right)\cos2\hat\phi_S\cos2\hat{\psi} \nonumber \\
        && \quad -\cos\hat\theta_S\sin2\hat\phi_S\sin2\hat{\psi} \,, \\ 
        \label{eqn:FxA}
        F^\times_A\left(\hat\theta_S,\hat\phi_S,\hat{\psi}\right) && =  
        \frac{1}{2}
        \left(1+\cos^2\hat\theta_S\right)\cos2\hat\phi_S\sin2\hat{\psi} \nonumber \\
        && \quad +\cos\hat\theta_S\sin2\hat\phi_S\cos2\hat{\psi} \,, \\
        \label{eqn:F+B}
        F^+_{B}\left(\hat\theta_S,\hat\phi_S,\hat{\psi}\right) && =  F^+_A(\hat\theta_S,\hat\phi_S-\pi/4,\hat{\psi})\,, \\
        \label{eqn:FxB}
        F^\times_{B}\left(\hat\theta_S,\hat\phi_S,\hat{\psi}\right) && = F^\times_A(\hat\theta_S,\hat\phi_S-\pi/4,\hat{\psi})\,.
    \end{eqnarray}
\end{subequations}
Here, a triangle GW detector like LISA can be equivalent to two L-shaped
detectors like LIGO denoted as detectors `A' and `B'. The ``hatted'' coordinate
implies it is tied to the LISA's detector frame while the ``unhatted'' one is
tied to the ecliptic frame. The unit vector of the GWs propagation direction
$\boldsymbol{N}$ and unit vector of CBCs orbital angle momentum $\boldsymbol{L}$
can be described by $(\theta_S,\,\phi_S)$ and $(\theta_L,\, \phi_L)$ in the
ecliptic frame, respectively. Following the method in Ref.~\cite{Cutler:1997ta},
we can relate $\left\{\hat\theta_S,\, \hat\phi_S,\, \hat{\psi}\right\}$ in
Eq.~\eqref{eqn:PF} to the ecliptic coordinates $\{\theta_S,\, \phi_S,\, \theta_L
,\, \phi_L\}$ through:

\begin{subequations}
    \label{eqn:PF_ecliptic}
        \begin{eqnarray}
             && \cos\hat\theta_S  = \frac{1}{2}\cos\theta_S-\frac{\sqrt{3}}{2}\sin\theta_S\cos\bar\phi_S \,, \\
             && \hat\phi_S = \frac{2\pi t}{T} - \frac{\pi}{12} + \tan^{-1}\left(\frac{\sqrt{3}\cos\theta_S+\sin\theta_S\cos\bar\phi_S}{2\sin\theta_S\sin\bar\phi_S}\right) \,,\\
             && \tan\hat{\psi} = \frac{\boldsymbol{\hat z\cdot p }}{\boldsymbol{\hat z\cdot q}}\,,\\
             && \boldsymbol{L} = \left(\sin\theta_L\cos\phi_L,\, \sin\theta_L\sin\phi_L,\, \cos\theta_L\right)\,,\\
            && \boldsymbol{N} = \left(\sin\theta_S\cos\phi_S,\, \sin\theta_S\sin\phi_S,\, \cos\theta_S\right)\,, \\
            && \boldsymbol{\hat z} = \left(-\frac{\sqrt{3}}{2}\cos\bar\phi,\,-\frac{\sqrt{3}}{2}\sin\bar\phi,\,\frac{1}{2} \right)\,,
    \end{eqnarray}
\end{subequations}
where $\boldsymbol{\hat z }$ is the unit vector in $\hat z$ direction,
$\bar\phi_S = \bar\phi - \phi_S$, and $\bar\phi = 2\pi t/T$ is the coordinate of
the detector at time $t$ with LISA orbit period $T = \rm 1 \, yr$. 

For a GW signal, we can obtain $t(f)$ from the GW phase evolution $\Phi(f)$ at
$0$PN oder approximately~\cite{Buonanno:2009zt},
\begin{eqnarray}
    \label{eqn:time_by_freq}
    t(f) = \frac{1}{2 \pi} \frac{\rmd \Phi(f)}{\rmd f} = t_c - 5(8\pi f )^{-8/3}\calM^{-5/3}\,.
\end{eqnarray}
Note that, this relation is derived from the inspiral stage with the stationary
phase approximation~\cite{Buonanno:2009zt}.  However, this is still a good
approximation for the entire stage of GW due to the rapid evolution during the
merger and ringdown stages.

Space-borne GW detector LISA moves along an orbit that lays on the Sun's equator
during the observation period, so there is a phase modulation in the signal, the
so-called Doppler phase $\phi_D$,
\begin{eqnarray}
    \label{eqn:phi_D}
    \phi_D(t) = 2\pi f R \sin\theta_S\cos \left[\bar\phi(t) - \phi_S \right]\,,
\end{eqnarray}
where $R$ = 1 au is the distance between LISA and the Sun. We can easily combine
$\phi_D(t)$ and
Eq.~\eqref{eqn:time_by_freq} to obtain $\phi_D(f)$ in Eq.~\eqref{eqn:hf}. In
addition, we can also obtain the inclination angle $\iota$ by 
\begin{eqnarray}
    \label{eqn:iota}
    \cos \iota = \boldsymbol{L}\cdot \boldsymbol{N}\,.
\end{eqnarray}

\subsection{Higher harmonic waveforms}
\label{subsec:waveform}

\begin{figure}[htp]
    \includegraphics[scale=0.5]{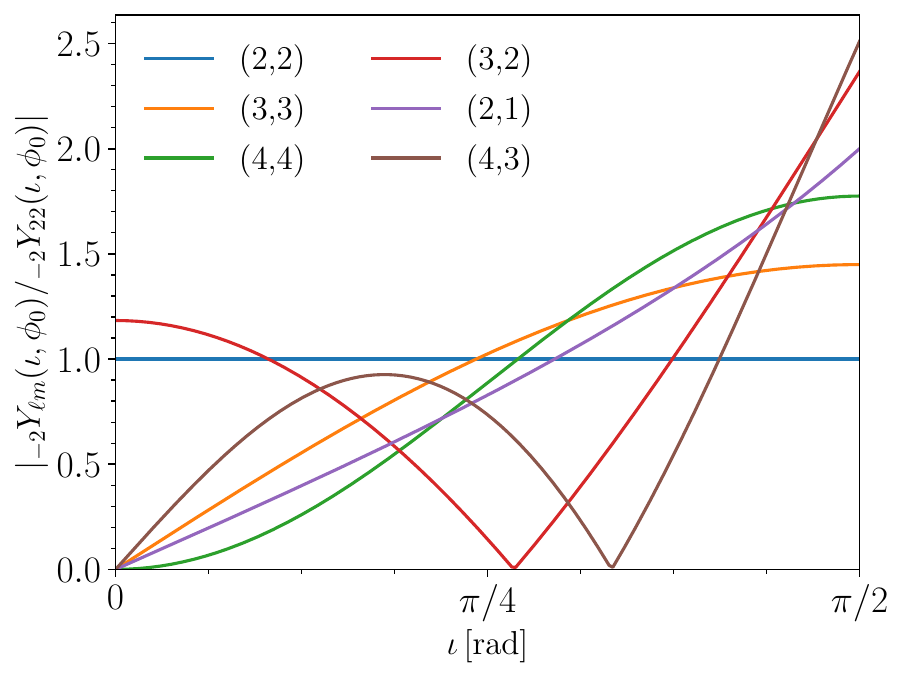}
\caption{\label{fig:spin_weighted} The amplitude ratio of the spin-weighted
spherical harmonic to ${}_{-2}Y_{2 2}(\iota,\phi_0)$ , where $\phi_0 = 0$. Each
line corresponds to a different harmonic mode.}
\end{figure}

The time domain `plus' and `cross' components $h_+(t)$ and $ h_\times(t)$ can be
decomposed into higher harmonics~\cite{Blanchet:2013haa,Buonanno:2006ui}:
\begin{eqnarray}
    \label{eqn:HM_time_domain}
     h_+(t) - \ii h_\times(t) = &&\sum_{\ell,m} {}_{-2}Y_{\ell m}(\iota,\, \phi_0) h_{\ell m}(t\,; \calM ,\, \eta,\, \DL,\, t_c)\,, \\
    \label{eqn:spin_weighted}
    {}_{-2}Y_{\ell m}(\iota,\phi_0) = &&\sqrt{\frac{2\ell+1}{4\pi}}d^{\ell m}_{-2}(\iota)e^{\ii m \phi_0} \,.
\end{eqnarray}
Here, ${}_{-2}Y_{\ell m}(\iota,\phi_0)$ is spin-weighted spherical harmonic
function with weight $-2$, and $d^{\ell m}_{-2}$ is a specific of Wigner
d-function, $\ell\ge 2$ and $|m|\le \ell$, leading to significant contributions
for different harmonics. Here, we illustrate the amplitude ratio of
${}_{-2}Y_{\ell m}(\iota,\phi_0)$ relative to that of $(\ell, m) = (2,2)$ in
Fig.~\ref{fig:spin_weighted}. When considering a face-on or face-off $(\iota = 0
,\, \pi)$ CBC, there are only two harmonics $(\ell, m) = (2,2)$ and $(3, 2)$
included in the waveform. For an edge-on $(\iota = \pi/2)$ system, the relative
contributions from higher harmonics are large.

The frequency domain waveforms $\tilde h_+(f)$ and $\tilde h_\times(f)$  can be
obtained from the time-domain GW strain by Fourier transform,
\begin{eqnarray}
    \label{eqn:Fourier_Transform}
    \tilde h_{+,\times}(f) = \int_{-\infty}^\infty  h_{+,\times}(t)e^{2\pi\ii f t} \rmd t \,.
\end{eqnarray}

Since $h_+(t)$ and $h_\times(t)$ are real functions of time, from
Eq.~\eqref{eqn:HM_time_domain} we can yield that
\begin{subequations}
    \label{eqn:h_timedomain_1}
    \begin{eqnarray}
        \label{eqn:hplus_t}
        h_+ &&= \frac{1}{2}\sum_{\ell,m} \left[ {}_{-2}Y_{\ell m}(\iota,\, \phi_0) h_{\ell m} + {}_{-2}Y^*_{\ell m}(\iota,\, \phi_0)h^*_{\ell m} \right]\,, \\
        \label{eqn:hcross_t}
        h_\times &&= \frac{\ii}{2}\sum_{\ell,m}  \left[{}_{-2}Y_{\ell m}(\iota,\, \phi_0) h_{\ell m} - {}_{-2}Y^*_{\ell m}(\iota,\, \phi_0)h^*_{\ell m}\right]\,. 
    \end{eqnarray}
\end{subequations}

For a non-precessing BBH, there exists a useful symmetry property for higher
harmonics: $h_{\ell m}^*(t) = (-1)^\ell h_{\ell,\, - m}(t)$. With the symmetry
property of time domain waveforms and performing Fourier transform, we can
obtain, 
\begin{eqnarray}
    \label{eqn:h_frequency_domain_simplify}
    \tilde h_{+,\, \times} (f)= \sum_{\ell m}\mathcal{Y}^{+,\, \times}_{\ell m}\tilde h_{\ell m}(f)\,,
\end{eqnarray}
where 
\begin{subequations}
    \begin{eqnarray}
        \mathcal{Y}^+_{\ell m} &= \frac{1}{2}\left[ {}_{-2}Y_{\ell m}(\iota,\, \phi_0)  + (-1)^{\ell}{}_{-2}Y^*_{\ell,\, -m}(\iota,\, \phi_0) \right]\,,\\
        \mathcal{Y}^\times_{\ell m} &= \frac{\ii}{2} \left[{}_{-2}Y_{\ell m}(\iota,\, \phi_0)  - (-1)^\ell {}_{-2}Y^*_{\ell,\, -m}(\iota,\, \phi_0)\right]\,.
\end{eqnarray}
\end{subequations}

Based on the discussion in Sec.~\ref{subsec:pattern functions} and this section,
9 parameters can characterize the GW signal in LISA. We list them in
Table~\ref{tab:parameter}, with $\mu_S\equiv \cos\theta_S$ and $\mu_L\equiv
\cos\theta_L$. The parameters' range means the value used in this investigation.
For convenience, we use the notation in Table~\ref{tab:HMnotation} to
indicate which harmonics are used.

\begin{table}
\caption{\label{tab:parameter}Parameters that characterize the GW signal. Here,
we fix the redshift $z=1$ and the luminosity distance $\DL=6791.27$ Mpc. The
direction of sources' angular momentum is fixed to be $\left(\mu_L,
\phi_L\right) = (0.3, 2.0)$. The time to coalescence $t_c = 0 $ and $\phi_0 =
0$.}
\begin{ruledtabular}
\begin{tabular}{ccc}
Notation  &Range  & Unit \\
\hline
$\mu_S$   & [$-1$,\, 1]     & 1 \\
$\mu_L$   & 0.3     & 1 \\
$\phi_S$  & [0,\, 2$\pi$] & rad \\
$\phi_L$  & 2.0 & rad \\
$\calM$   & $\left[\calM_{\rm min},\, \calM_{\rm max}\right]$ & $\SUN$ \\
$\eta$    & $\left[\eta_{\rm min},\,\eta_{\rm max} \right]$   &1\\
$\DL$     &6791.27        & Mpc\\
$t_c$     &0              & s\\
$\phi_0$  &0              & rad
\end{tabular}
\end{ruledtabular}
\end{table}

\begin{table}
\caption{\label{tab:HMnotation}Different scenarios for investigating the effects
from higher harmonics in data analysis for LISA.}
\begin{ruledtabular}
\begin{tabular}{ll}
Notation &($\ell$,$|m|$) \\
\hline
 $\rm{I}$ & (2,2)  \\
 $\rm{II}$ & (2,2),\, (3,3)\\
 $\rm{III}$ & (2,2),\, (3,3),\, (4,4)\\
 $\rm{IV}$ & (2,2),\, (3,3),\, (4,4),\, (3,2)\\
 $\rm{V}$ & (2,2),\, (3,3),\, (4,4),\, (3,2),\, (2,1)\\
 $\rm{VI}$ & (2,2),\, (3,3),\, (4,4),\, (3,2),\, (2,1),\, (4,3)
\end{tabular}
\end{ruledtabular}
\end{table}

\begin{figure}[htp]
    \includegraphics[scale=0.55]{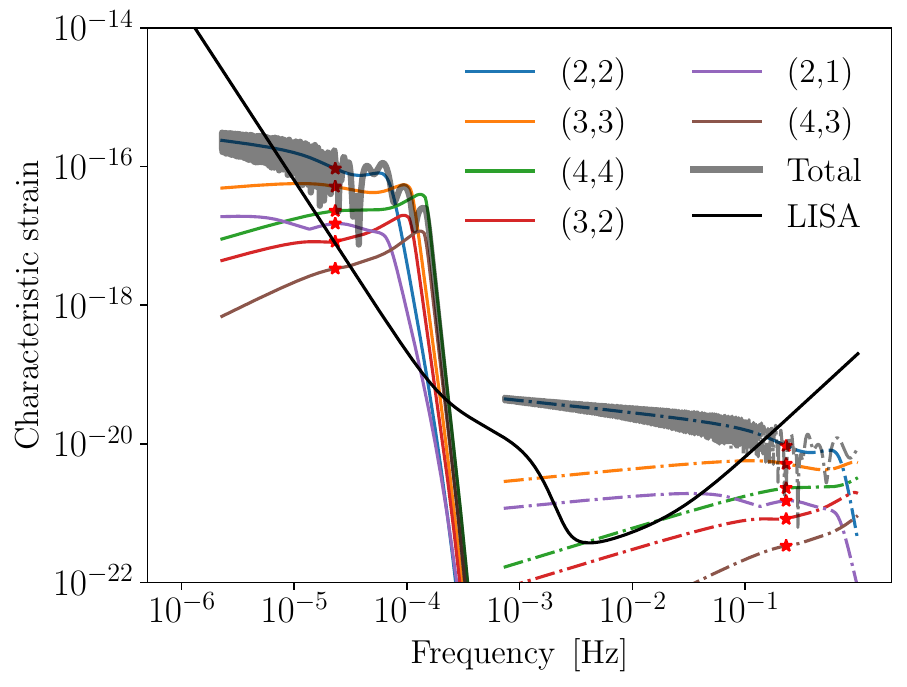}
\caption{\label{fig:charactieristic_strain} The characteristic strains $2f
\left|\tilde{h}_+(f) \right|$ of different BBHs in LISA 4-yr observation. We set their
physical parameters $q =10,\, \iota =\pi/2,\,z =1$ with total mass $M =
10^8\,\SUN$ (solid lines) and $M = 10^4\,\SUN$ (dashed lines).  The solid black
line in the figure is the characteristic strain of the detector noise, $\sqrt{f
S_n(f)}$. Each colored line represents a single harmonic component of GWs. Red
stars denote the innermost stable circular orbit (ISCO) frequency, and the thick
gray line is the total strain of all harmonics in the figure.}
\end{figure}

We investigate the characteristic strain for the two BBH systems and compare the
individual harmonics strain and the total strain.
Figure~\ref{fig:charactieristic_strain} shows the each harmonics of GWs with $M
= 10^4\,\SUN$ and $M = 10^8\,\SUN$. We can find that when $M$ is relatively
small, only $(3,3)$ and $(4,4)$ modes are significant. But when $M$ is
relatively large, more higher-order modes are visible, and the effect may not be
neglected. This is mainly because the higher-order modes contribute more to the
strain amplitude in the merger and ringdown stages. For the space-borne GW
detector LISA, MBH binary systems need to include higher modes to extract more
information from the observed data.

\subsection{SNRs}
\label{subsec:SNR}

We adopt the matched-filtering method~\cite{Finn:1992wt} to estimate the SNR of
GWs. For a given GW strain $\tilde{h}(f)$, the optimal SNR is defined as,
\begin{eqnarray}
    \label{eqn:SNR}
    \rho \equiv \left(\tilde{h}(f) \right|\left.\tilde{h}(f) \right)^{1/2}\,,
\end{eqnarray}
where operator $\left(\tilde{\mathcal{A}}\right|\left.\tilde{\mathcal{B}}
\right)$ is the noise-weighted inner product between two signals
$\tilde{\mathcal{A}}(f)$ and $\tilde{\mathcal{B}}(f)$,
\begin{eqnarray}
    \label{eqn:inner_product}
    \left(\tilde{\mathcal{A}}\right|\left.\tilde{\mathcal{B}} \right) = 2\int_{f_{\rm low}}^{f_{\rm high}}\frac{\tilde{\mathcal{A}}(f)\tilde{\mathcal{B}}^*(f)+\tilde{\mathcal{A}}^*(f)\tilde{\mathcal{B}}(f)}{S_n(f)}\rmd f \,.
\end{eqnarray}
In this paper, we set $f_{\rm high}=1 \,{\rm Hz}$ for LISA and deduce $f_{\rm low}$ by,
\begin{eqnarray}
    \label{eqn:f_low}
    f_{\rm low}= \left(\frac{(8\pi)^{8/3}T_{\rm obs}}{5\calM^{5/3}} -f_{\rm high}^{-8/3}\right)^{-3/8}\,.
\end{eqnarray}
$S_n(f)$ is the one-sided power spectrum density of noise. We adopt the LISA
sensitivity curves in Ref.~\cite{Robson:2018ifk}, with $L = 2.5\times 10^9 \
{\rm m}$ as the arm length of LISA, and an observation span $T_{\rm obs} = {\rm
4 \, yr}$.

\begin{figure}[htp]
    \centering

    \includegraphics[scale=0.55]{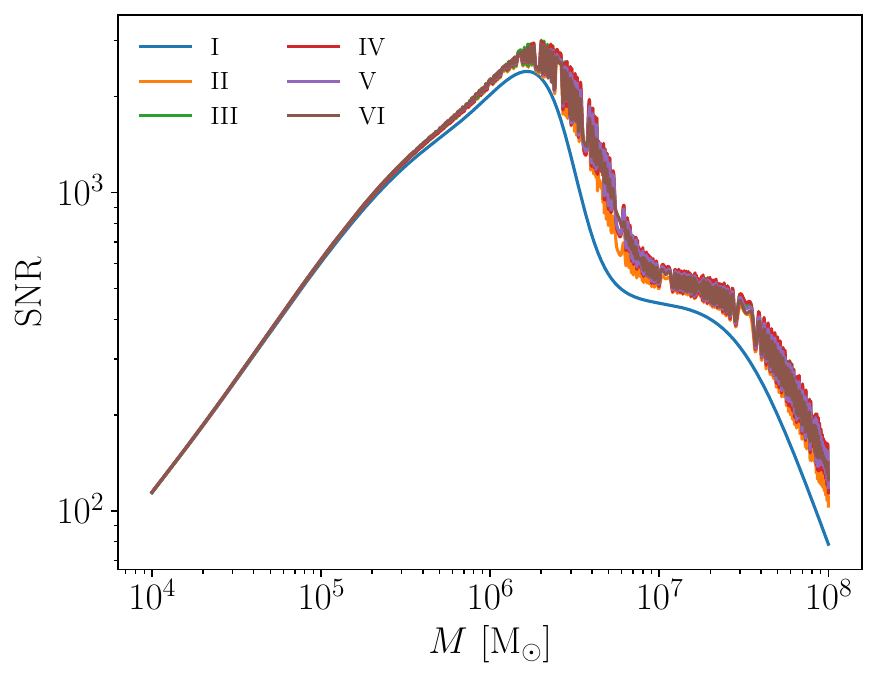}
    
\caption{The SNRs as a function of total mass $M$. The sources are located at $z =
1$, mass ratio $q = 10$, $\phi_0 = 0$, $t_c = 0\,\rm{s}$, and angular parameters
$\left( \mu_S, \phi_S, \mu_L, \phi_L \right) = (-0.25, 2.31, 0.3, 2.0)$.  Each
curve corresponds to a different scenario and adopts the same notation as
Table~\ref{tab:HMnotation}. Since the SNRs of the different scenarios are close,
several curves overlap. With the increasing $M$, the oscillation increases, which
means the effect of higher harmonics increases.  \label{fig:snr_total_mass}}
\end{figure}

\begin{figure}[htp]
    \centering

    \includegraphics[scale=0.55]{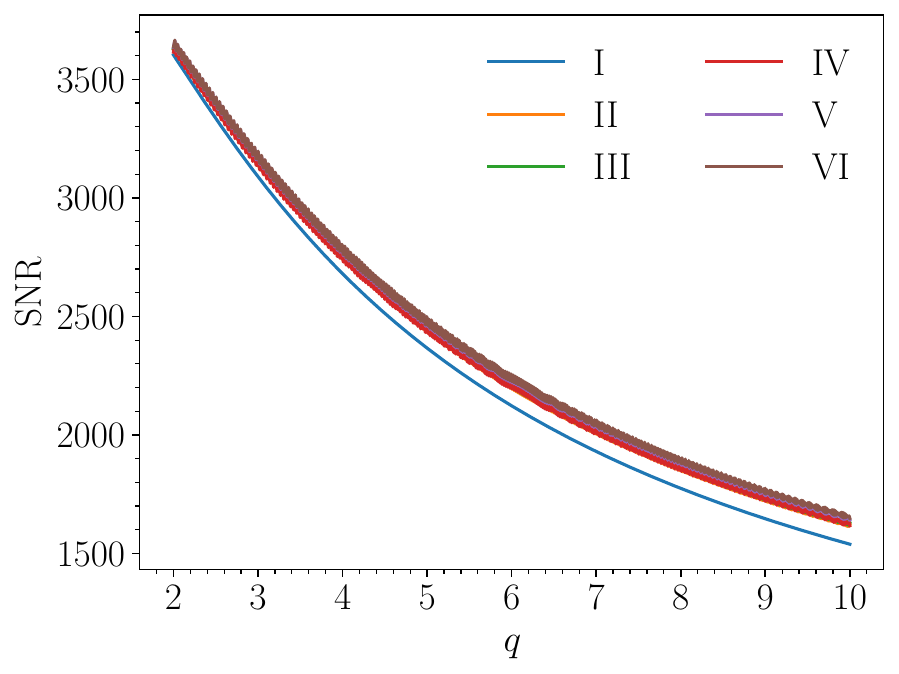}
\caption{ The SNRs as a function of mass ratio $q$. The sources' parameters are
the same as Fig.~\ref{fig:snr_total_mass} except total mass $M = 1.1\times
10^6\, \SUN.$ Each curve corresponds to a different scenario and adopts the same
notation as Fig.~\ref{fig:snr_total_mass}. Since we adopt a system with total
mass $M = 1.1\times 10^6\, \SUN,$ the effect of higher harmonics modes is
significant (c.f. Fig.~\ref{fig:charactieristic_strain}), so the oscillations of
SNRs appear when incorporating higher harmonics.  \label{fig:snr_mass_ratio}}
\end{figure}

From Fig.~\ref{fig:charactieristic_strain}, we show the effect of higher
harmonics relating to the total mass $M$. The corresponding SNRs are illustrated
in Fig.~\ref{fig:snr_total_mass}, which the sources located at $z =  1$, the
total mass $M$ varies from $10^4\,\SUN$ to $10^8\, \SUN$, and the other
parameters of the sources are $q = 10,\,\mu_S = -0.25,\,\phi_S = 2.31,\,\mu_L =
0.3,\,\phi_L = 2.0,\,\phi_0 = 0$ and $t_c = 0\,\rm{s}$. The deviations between
$(2,2)$ mode and other scenarios increase as $M$ increases, as implied in
Fig.~\ref{fig:charactieristic_strain}. Meanwhile, Our analysis indicates that
LISA is most sensitive to $M \sim 10^6\, \SUN$, and SNRs exhibit oscillatory
behavior as $M$ increases. This is due to SNRs being the inner product of $h(f)$
with itself, where $h(f)$ is a linear superposition of each harmonic. The cross
term of different harmonics can cause constructive or destructive interference,
which has been noted by \citet{Marsat:2020rtl,Mills:2020thr}. So in
Fig.~\ref{fig:snr_total_mass}, the SNRs curve is smooth when only involving
$(2,2)$ mode and oscillations appear when higher harmonics are involved. And for
the relatively small mass source, compared with $(2,2)$ mode, the effect of
higher harmonics is not so significant (see
Fig.~\ref{fig:charactieristic_strain}), therefore, the oscillatory behavior is
not notable for the scenarios that incorporate higher harmonics.

Besides, we also investigate the effect of varying mass ratio $q$ in
Fig.~\ref{fig:snr_mass_ratio}. The sources' parameters are the same as
Fig.~\ref{fig:snr_total_mass} except total mass $M = 1.1\times 10^6\, \SUN$, and
the mass ratio $q$ varies from $2$ to $10$. The oscillations also result from
constructive or destructive interference of different harmonics, the same as
Fig.~\ref{fig:snr_total_mass}. As expected, the SNR deviation between different
scenarios becomes more significant as $q$ increases, as the BBH becomes more
asymmetric. When the total mass is fixed, a larger $q$ leads to a smaller SNR,
which is related to a smaller $\calM$. In short words,
Figures.~\ref{fig:snr_total_mass} and~\ref{fig:snr_mass_ratio} imply that higher
harmonics are more significant for a heavier and asymmetric BBH. Moreover, the
oscillations that appear in Figs.~\ref{fig:snr_total_mass}
and~\ref{fig:snr_mass_ratio} will reappear in the subsequent parameter
estimation for the same reason.

\begin{figure}[htp]
    \includegraphics[scale=0.42]{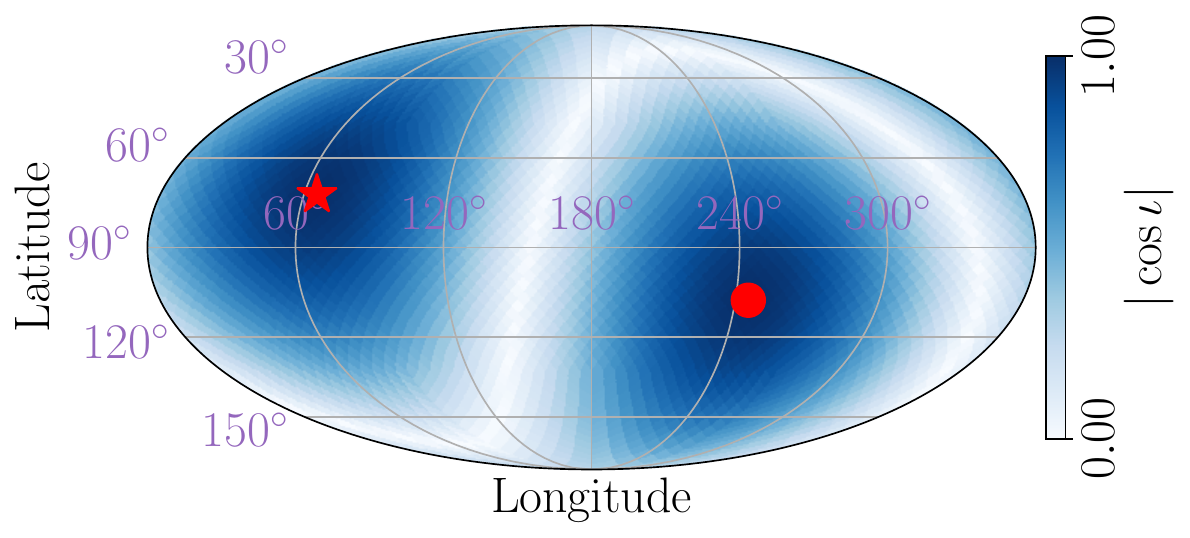}
\caption{\label{fig:iota}The sky map of $|\cos\iota|$ which $\mu_L = 0.3,\,
\phi_L = 2.0$.  The red star denotes $\cos\iota = 1$, i.e. the face-on case, and
the red circle denotes $\cos\iota = -1$, i.e. the face-off case. We can see the
edge-on cases as the white curve.}
\end{figure}

\begin{figure}[htp]
    \centering  
    {
        \includegraphics[scale=0.42]{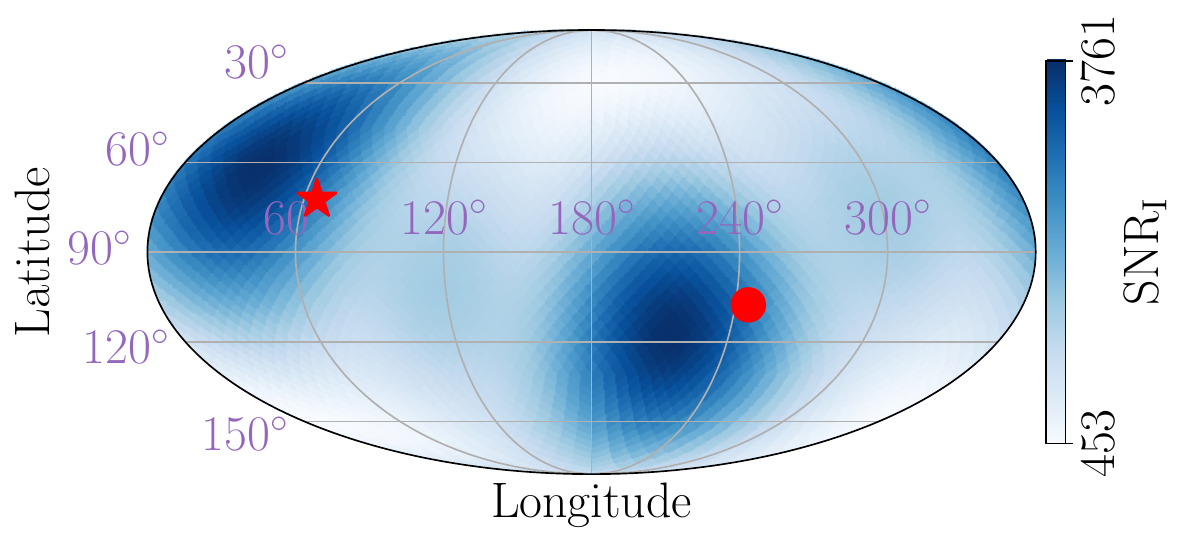}
} \caption{The SNRs sky map of a binary with $m_1 = 10^6\,\SUN$, $m_2 =
10^5\,\SUN$, $z=1$, and angular parameters $\left( \mu_S, \phi_S, \mu_L, \phi_L
\right) = (-0.25, 2.31, 0.3, 2.0)$, only $(\ell,|m|) = (2,2)$ mode is involved.
The red star and red circle denote the face-on case and face-off case,
respectively.  \label{fig:SNRskymap}}
\end{figure}

In Fig.~\ref{fig:spin_weighted}, we show that the higher harmonics strongly
depend on the inclination angle $\iota$, i.e. the BBH orbital angular momentum
and location. Thus, investigating the effect of location is promising. To
visualize the inclination angle $\iota$, we fix $\mu_L = 0.3,\, \phi_L =  2.0$
in this paper, let $\mu_S$ varies from $-1$ to $1$, and $\phi_S$ varies from $0$
to $2\pi$. The $|\cos\iota|$ is presented in Fig.~\ref{fig:iota}. We denote
$\cos\iota = 1$  as the face-on case, $\cos\iota = -1$  as the face-off case,
and $\cos\iota = 0$ is the edge-on case.  Then we investigate the effect of
sources' location and higher harmonics on SNRs,  which are shown in
Fig.~\ref{fig:SNRskymap}.

Moreover, we show the effect of source location $\left(\mu_S,
\phi_S\right)$ on SNRs in Fig.~\ref{fig:SNRskymap} as well. The sources are still located at $z =
1$, with $m_1 = 10^6\,\SUN,\,m_2 = 10^5\,\SUN,\, \mu_L =  0.3, \phi_L = 2.0,\,
\phi_c = 0,\, t_c  = 0\rm{s}$, $\mu_S$ varies from $-1$ to 1, and $\phi_S$
varies from $0$ to $2\pi$ (all information has been shown in Table~\ref{tab:parameter}). Here, we
only show the SNR results from the scenario that only includes the (2,2) mode.
The SNRs for this particular scenario are approximately $10^2\sim10^3$. Note
that, the loudest area is not the face-on nor face-off areas, but slightly
offset. When we consider a face-on or face-off binary system, indeed,
$\tilde{h}_+(f)$ and $\tilde{h}_\times(f)$ reach their maximum values. However,
it does not mean that $\tilde{h}(f)$ reaches its maximum values as well because
of the different pattern functions. Furthermore, the SNRs in the edge-on area
are much smaller, only about 10\% of the loudest areas.  This substantial SNR
difference will lead to better parameter extraction near the face-on and
face-off areas than at the edge-on areas, where the higher modes are prominent.
Additionally, we have also investigated the results with different binary
masses, which are consistent with the behaviors in Figure 6, and we do not
show them here.

\subsection{Parameter estimation}
\label{subsec:ParemeterEstimate}

Bayesian inference is used to acquire the posterior distribution of the source
parameters in GW transient data under the assumption that noise is Gaussian and
stationary~\cite{Veitch:2014wba,LIGOScientific:2018mvr,LIGOScientific:2020ibl,LIGOScientific:2021usb,LIGOScientific:2021djp}.
In principle, we should perform massive computational inference to obtain the
posterior distribution of the binary parameters to investigate the higher
harmonic effects. However, it seems to be improper to perform such a task.
Thanks to the large SNR of most GW signals we considered,\footnote{Note that,
the Fisher matrix method is only valid in the linear signal approximation (or
high-SNR approximation). More details can be found in
Ref.~\cite{Vallisneri:2007ev,Zhao:2021bjw}.} the sufficient and computationally
cheap method, Fisher information matrix~\cite{Finn:1992wt,Cutler:1994ys}, can be
easily employed to estimate the parameters of the sources.

To demonstrate how to use the Fisher matrix, we use $\boldsymbol{\mathcal{P}}$
to denote the set of these parameters, and $\hat{\boldsymbol{\mathcal{P}}}$ to
denote the parameters set that maximizes the likelihood function $\Lambda$,
where
\begin{eqnarray}
    \label{eqn:likelihood}
    \Lambda = {\rm{exp}} \left[\left(\tilde d(f)\right|\left.\tilde h(f)\right) - \frac{1}{2}\left(\tilde h(f)\right|\left.\tilde h(f)\right)\right]\,,
\end{eqnarray}
where $\tilde d(f)$ is the Fourier transform result of $d(t)$ and $d(t)$ is the
strain detected in the detector. Given the strain $d(t)$, the probability
density function of the physical parameters $\boldsymbol{\mathcal{P}}$
is~\cite{Finn:1992wt,Cutler:1994ys}
\begin{eqnarray}
    \label{eqn:fisher_probability}
    p(\boldsymbol{\mathcal{P}}|d) \propto {\rm{exp}}\left[  -\frac{1}{2}\sum_{i,j}\Gamma_{ij} \Delta{\mathcal{P}^i}\Delta{\mathcal{P}^j}\right] \,, 
\end{eqnarray}
where $\Gamma_{ij}$ is the Fisher information matrix (the inverse of the
covariance matrix) of
$\Delta\boldsymbol{\mathcal{P}} =\boldsymbol{\mathcal{P}} -
\hat{\boldsymbol{\mathcal{P}}} $. The elements
of
$\Gamma_{ij}$ can be obtained by,
\begin{eqnarray}
    \label{eqn:fisher_matrix}
    \Gamma_{ij} = \left(\frac{\partial \tilde h(f\,; \boldsymbol{\mathcal{P}})}{\partial {\mathcal{P}^i}}\right|\left.\frac{\partial \tilde h(f\,; \boldsymbol{\mathcal{P}})}{\partial {\mathcal{P}^j}}\right) \,.
\end{eqnarray}
For the detector LISA, there are two equivalent L-shaped detectors `A' and `B',
so we can obtain the final Fisher matrix by linear superimposing the individual
Fisher matrix,
\begin{eqnarray}
    \Gamma_{ij} = \Gamma^A_{ij}+\Gamma^B_{ij}\,.
\end{eqnarray}

For the parameters $\left\{\ln \DL,\, t_c,\, \phi_0\right\}$, we can yield the
analytic partial derivatives of them directly,
\begin{eqnarray}
    \frac{\partial \tilde h}{\partial \ln  \DL} &&= - \tilde h\label{eqn:dh_dDl}\,,\\
    \frac{\partial \tilde h}{\partial t_c}  &&= \ii 2\pi f\tilde h\label{eqn:dh_dtc}\,,\\
    \frac{\partial \tilde h}{\partial \phi_0} &&= \ii \sum_{\ell,m} m\left (F^+\mathcal{Y}^+_{\ell m} + F^\times \mathcal{Y}^\times_{\ell m}\right ) \tilde h_{ \ell m} e^{-\ii \phi_D}\,.
    \label{eqn:dh_dphi}
\end{eqnarray}
For the angular parameters, $\left\{\mu_S,\, \mu_L,\, \phi_S,\, \phi_L \right\}$,
we can use the chain rule to obtain their analytic partial derivatives. However,
their derivatives are too lengthy and uninspiring, so we do not show them here. 

For the parameters $\calM$ and $\eta$, we use central
difference method to gain their derivatives,
\begin{eqnarray}
\label{eqn:central_diff}
\frac{\partial \tilde h(f\,; \boldsymbol{\mathcal{P}})}{\partial {\mathcal{P}}^i} \simeq \frac{\tilde h(f\,; \boldsymbol{\mathcal{P}}+\delta {\mathcal{P}}^i)-\tilde{h}(f\,; \boldsymbol{\mathcal{P}}-\delta {\mathcal{P}}^i)}{2\delta {\mathcal{P}}^i}\,.
\end{eqnarray}

To estimate the numerical error, we compare the result from the central
difference method with that from \textsc{JAX}~\cite{JAX:2018Github}, a
\textsc{python} package performing the high-precision automatic differentiation
algorithm to obtain the derivatives. The relative errors between our numerical
method and \textsc{JAX} are under 1\%.

\section{Result and discussion}
\label{sec:Result}

In this paper, we focus on the higher-harmonic effect on the parameter
estimations of the chirp mass, symmetric mass ratio, luminosity distance, and
sky location. Besides, we also investigate how the total mass $M$, mass ratio
$q$, and source location impact the parameter estimations.

The angular resolution $\Delta\Omega$ can
be expressed as~\cite{Cutler:1997ta,Barack:2003fp},
\begin{eqnarray}
\label{eqn:skylocation}
\Delta\Omega = 2\pi\sqrt{\left(\Delta\mu_S  \Delta \phi_S\right)^2 - \left[\left(\Gamma^{-1}\right)_{\mu_S\phi_S}\right]^2}\,.
\end{eqnarray}

To make the results organized, we present effects from the total mass $M$ and
the mass ratio $q$ in Sec.~\ref{subsection:mass}, and the impact of the sources'
location in Sec.~\ref{subsection:skymap}.

\subsection{The effect of total mass and mass ratio}
\label{subsection:mass}

\begin{figure*}[htp]
    \centering

    \includegraphics[scale=0.55]{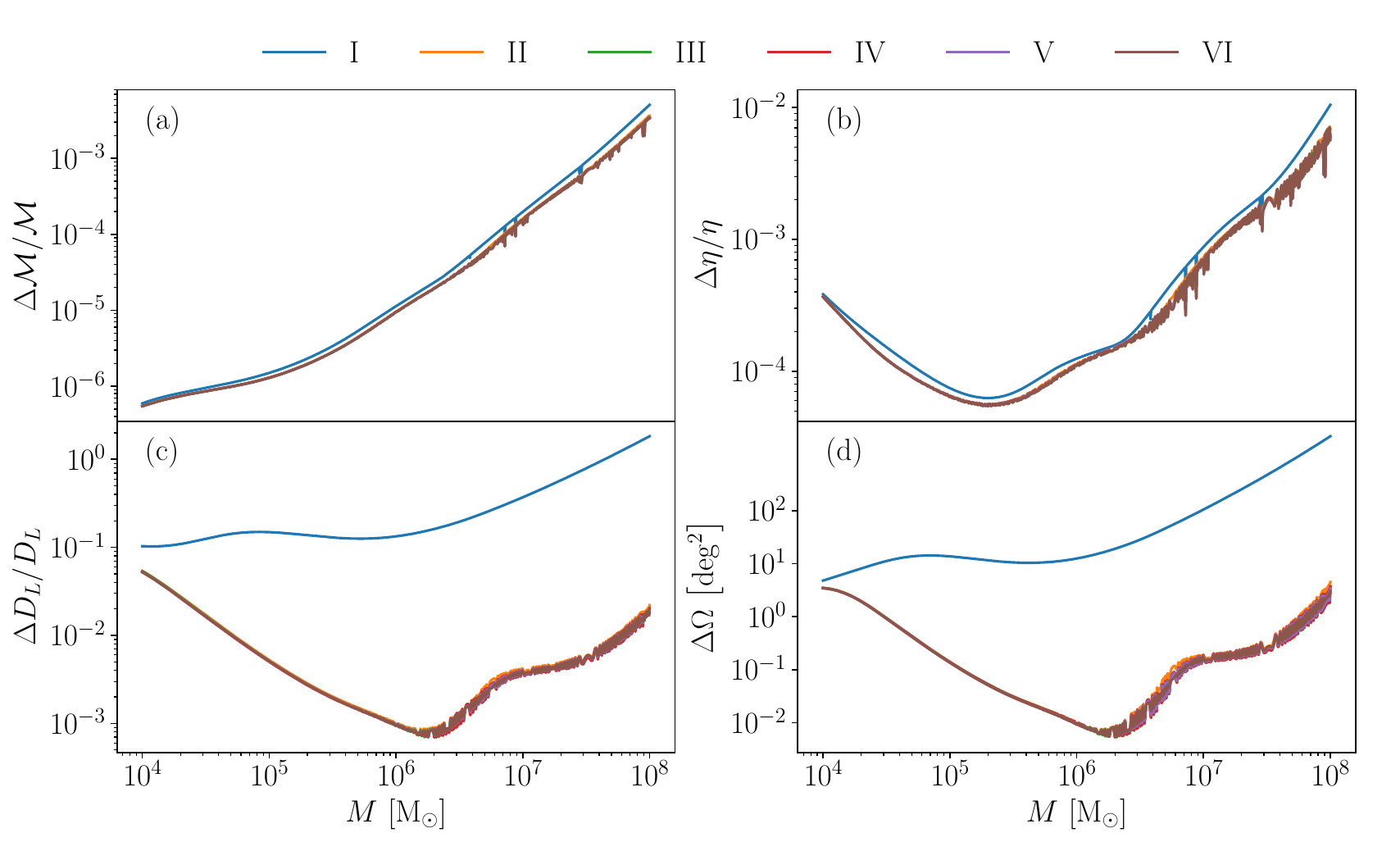}
    
\caption{The constraints on the chirp mass [panel (a)], the symmetric mass ratio
[panel (b)], the luminosity distance [panel (c)], and the angular resolution
[panel (d)] with the varying total mass $M$. We fix the mass ratio $q=10$ and
angular parameters $\left( \mu_S, \phi_S, \mu_L, \phi_L \right) = (-0.25, 2.31,
0.3, 2.0)$. Each curve represents a different scenario in
Table.~\ref{tab:HMnotation}.  \label{fig:total_mass}}.
\end{figure*}

\begin{figure*}[htp]
    \centering

    \includegraphics[scale=0.55]{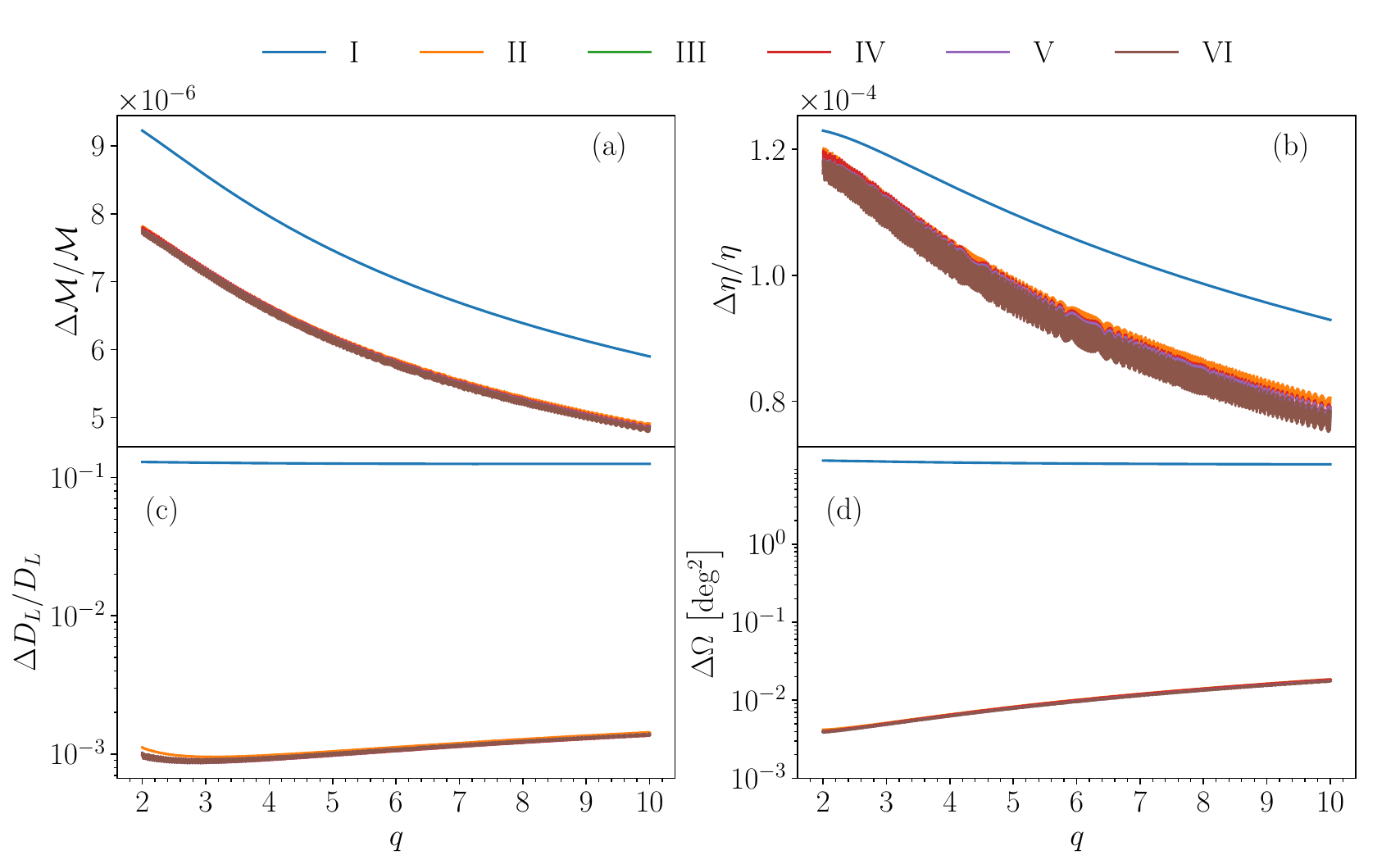}
\caption{Same as Fig.~\ref{fig:total_mass} with the varying mass ratio $q$. We
fix the total mass $1.1 \times 10^{6}\,{\rm \SUN}$.
    \label{fig:mass_ratio}}
\end{figure*}

The effects of varying total mass $M$ and higher harmonics impact parameter
estimation are illustrated in Fig.~\ref{fig:total_mass}, which take the same
sources' parameters as in Fig.~\ref{fig:snr_total_mass}. There exists an
oscillatory behavior of SNRs (see Fig.~\ref{fig:snr_total_mass}). Such
oscillation is caused by the interference of each harmonic. In
Fig.~\ref{fig:total_mass}, a similar phenomenon reappears. Because the
parameters' variances are the diagonal elements of the
$\left(\Gamma^{-1}\right)_{ij}$, and $\Gamma_{ij}$ are the inner product of
waveform $\tilde{h}(f)$ derivatives. Meanwhile, the derivatives can be regarded
as a linear superposition of each harmonic derivative since the $\tilde{h}(f)$
can be regarded as a linear superposition of each harmonic, which will lead to
parameter extraction oscillation as $M$ increases.

When higher harmonics are involved, the precisions on $\calM$ and $\eta$
increase by factors of $\sim 2$ and $\sim 3$ at most, respectively. As expected,
the effect of higher harmonics becomes visible for $\calM$ and $\eta$ resolution
with increasing $M$. The angular and $\DL$ resolutions increase by factors of
$\sim 3000$ and $\sim 300$ at most, respectively. Refer to
Fig.~\ref{fig:snr_total_mass}, roughly speaking, the effect of higher harmonics
beyond $(3,3)$ can be neglected in this case, and a higher SNR means higher
parameter estimation precision. Although Fig.~\ref{fig:charactieristic_strain}
implies the higher harmonics are more significant for the heavier BBH. Our
result shows the
significant improvement for $\DL$ and $\Omega$, as
shown in Fig.~\ref{fig:total_mass} appears at $M\sim 10^6\,\SUN$, where their
merger frequencies correspond
to the most sensitive frequency $\sim 2\times10^{-3}\,{\rm Hz}$ for LISA. 

In Fig.~\ref{fig:total_mass}, we notice whether or not including higher modes
into data analysis, the $\calM$, and $\eta$ estimations are highly precise.
This is because the parameters $\calM$ and $\eta$ mainly contribute to the GW
phase $\Phi(f)$, which can be detected at a highly precise level with a
long-lived signal. As the $M$ increases, the inspiral of MBH binary becomes
shorter, resulting in a decrease in precisions on $\calM$, and $\eta$.  In
addition, the estimations of coalescence time $t_c$ and $\phi_0$ are also at a
highly precise level for the same reason. However, for the inference of $\DL$
and $\Omega$, we mainly rely on two effects: one is the Doppler effect
$\phi_D\left(\theta_S,\phi_S\right)$ [see Eq.~\eqref{eqn:phi_D}]; the other one
is the LISA pattern factor functions
$F^{+,\times}\left(\hat\theta_S,\hat\phi_S,\hat{\psi}\right)$ [see
Eq.~\eqref{eqn:PF}]. When the merger frequencies of MBH systems are in the most
sensitive band of LISA, these GW features with higher modes will be well
monitored by LISA. The interference of higher harmonics will be significant and
easily be extracted from the data, resulting in breaking the $\DL$-$\iota$
degeneracy, [see Eqs.~\eqref{eqn:HM_time_domain} and~\eqref{eqn:spin_weighted}].
and reach a highly precise level. 

Therefore, in Fig.~\ref{fig:total_mass}, we will find that including higher
harmonics only slightly affects $\calM$ and $\eta$ estimations but significantly
impact $\DL$ and $\Omega$ resolutions. Furthermore, it implies that the impact of
including higher harmonics, for $\DL$ and $\Omega$ resolution, is positively
correlated to SNRs. This is the result that, compared with higher harmonics
itself, the higher harmonics effect is more significant with the larger SNRs. 

We also show the effects of varying mass ratio $q$ and higher harmonics in
Fig.~\ref{fig:mass_ratio}, which takes the same sources' parameters as in
Fig.~\ref{fig:snr_mass_ratio}. Similar to Fig.~\ref{fig:total_mass}, whether or
not including higher modes into data analysis, the estimations for $\calM$ and
$\eta$ estimations are similar for the different scenarios, which at the order
$10^{-5}$ and $10^{-4}$, respectively. However, the resolutions of $\DL$ and
$\Omega$ are at the order of 10\% and 10 $\rm deg^2$ precisions with only
$(2,2)$ mode, respectively. When including higher modes, even with only $(3,3)$
mode, the estimations for $\DL$ and $\Omega$ are improved massively. The results
are consistent with the studies for varying $M$.

Refer to Fig.~\ref{fig:snr_mass_ratio}, the oscillation is significant, and a
similar phenomenon reappears. Fig.~\ref{fig:mass_ratio} shows the increases of
$\calM$ and $\eta$ resolutions are tiny (a factor of $\sim 1.2$ at most) when
incorporating the higher harmonics. The angular and $\DL$ resolutions increase
dramatically by factors of $\sim 10^3$ and $\sim 10^2$ at most respectively.
Meanwhile, the increase in $q$ leads to less improvement for the angular and
$\DL$ resolution since SNRs decrease. In addition to the $(3,3)$ mode, we find
that the $(4,4)$ mode also has a slight impact on parameter extraction [see
Fig.~\ref{fig:mass_ratio} (b)]. This also is implied in
Fig.~\ref{fig:total_mass} but is not significant.

In short words, the higher harmonics slightly affect $\calM$ and $\eta$
resolutions but impact dramatically on $\DL$ and $\Omega$ resolutions. And the
larger SNRs mean a better $\DL$ and $\Omega$ resolution improvement. This is to
be expected as we explained for Fig.~\ref{fig:total_mass}.

In addition, \citet{Baibhav:2020tma} found that the errors of $\DL$ and $\iota$
will diverge at $q\simeq 4$. This is because they used the analytic MBH ringdown
waveform to estimate the parameters, whereas we use the full waveform.

\subsection{The effect of source location}
\label{subsection:skymap}

\begin{figure*}[htp]
    \centering
 
    \includegraphics[scale=0.28]{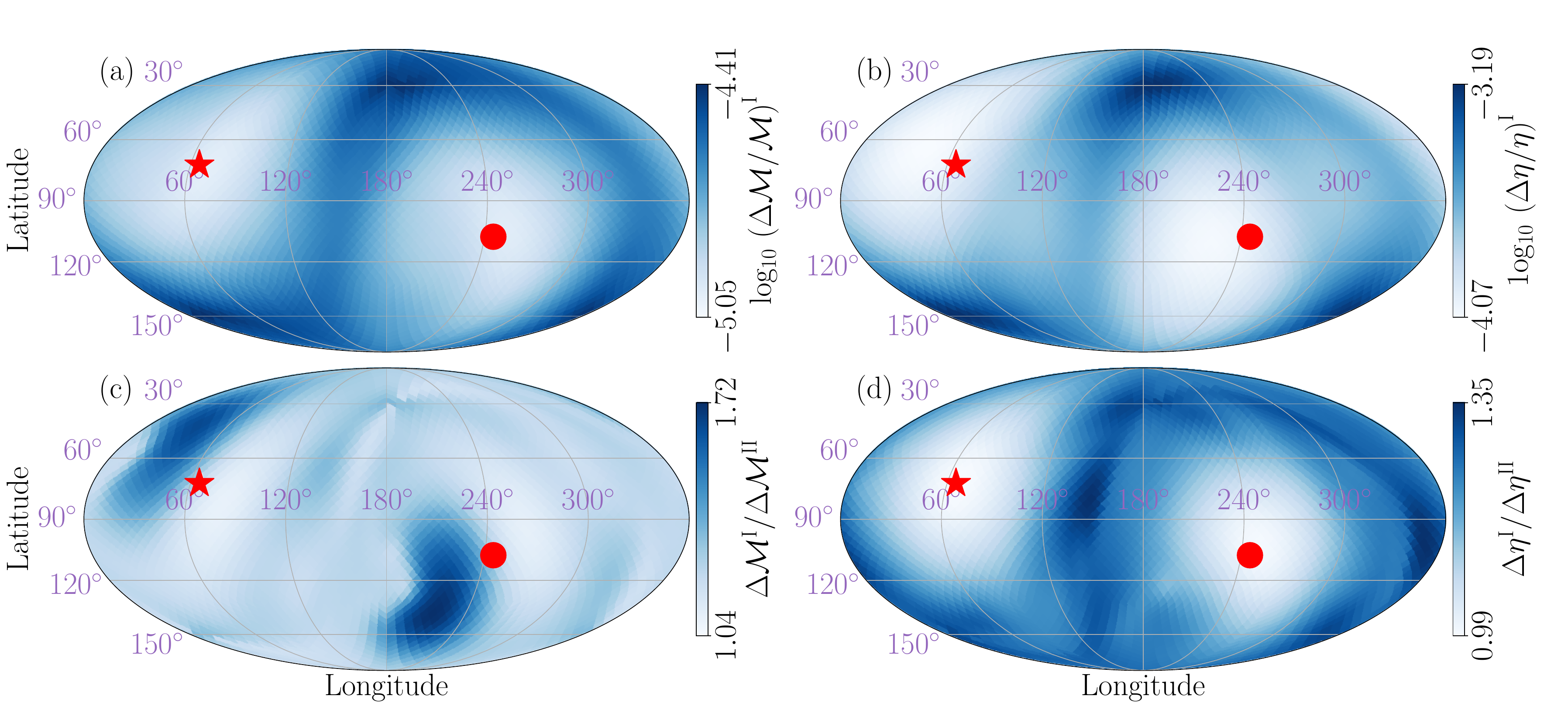}
\caption{The skymaps of $\calM$ and $\eta$ resolutions for the binary systems
with the same source parameters as Fig.~\ref{fig:SNRskymap}. The superscripts
`I' and `II' denote the different scenarios in Table~\ref{tab:HMnotation}. The
top panel shows $\rm{log_{10}}\, \left(\Delta\calM/\calM\right)^ {\rm I}$ and
$\rm{log_{10}}\, \left(\Delta\eta/\eta\right)^{\rm I}$ which are only
incorporating the $(2,2)$ harmonic mode, and the bottom panel shows
$\Delta\calM^{\rm {I}}/\Delta\calM^{\rm {II}}$ and $\Delta\eta^{\rm
{I}}/\Delta\eta^{\rm {II}}$, the ratio of scenario I to scenario II for $\calM$
and $\eta$ resolutions. The red star and red circle denote the face-on case and
face-off case, respectively.  
    \label{fig:calMetaSkymap}}
\end{figure*}

\begin{figure*}[htp]
    \centering
    \includegraphics[scale=0.28]{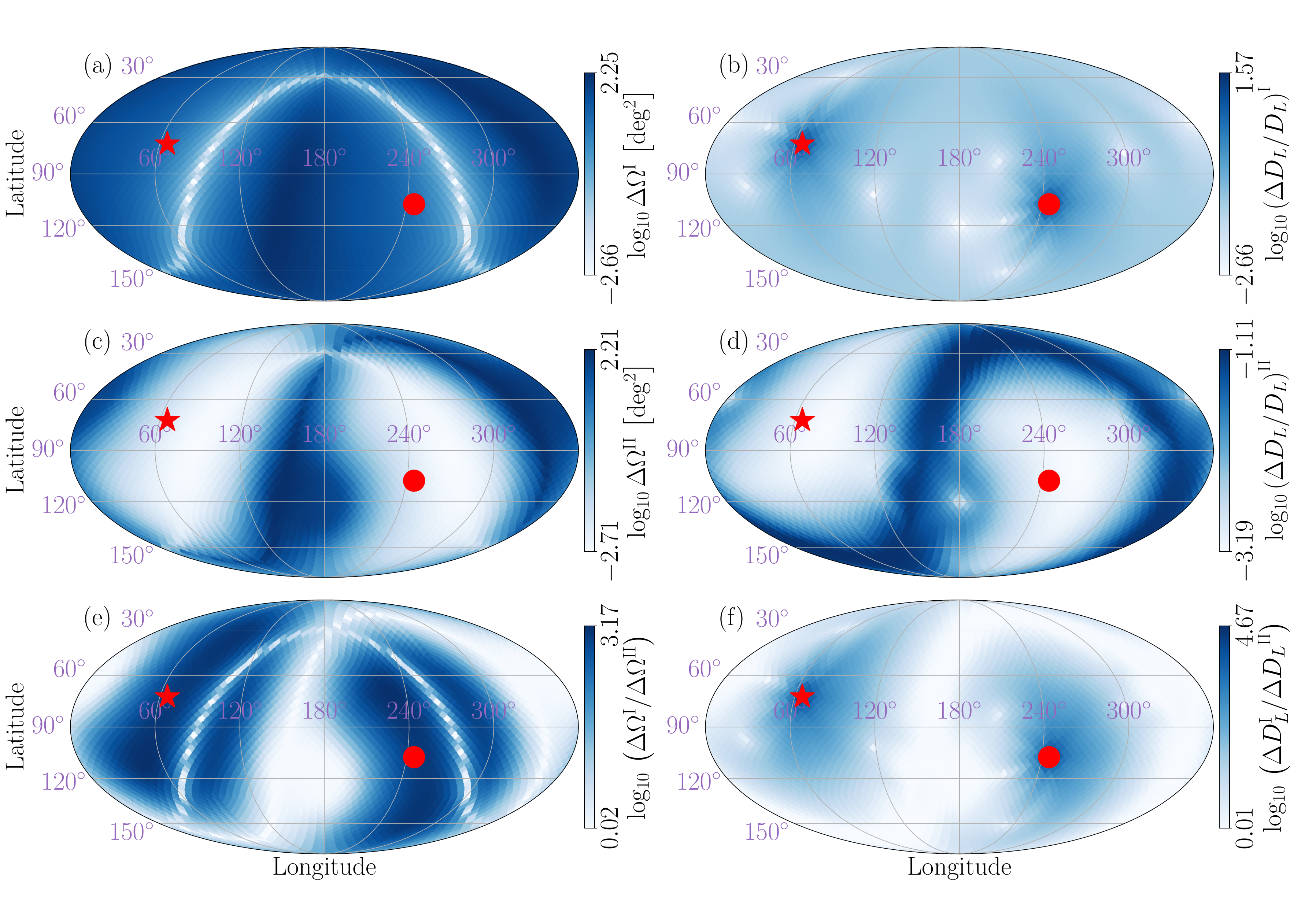}
\caption{Same as Fig.~\ref{fig:calMetaSkymap}, but for angular and luminosity
distance resolution, $\Omega$ and $\DL$. The top panel shows $\rm{log_{10}}\,
\Delta\Omega^I$ and $\rm{log_{10}}\, \left(\Delta\DL/\DL\right)^{\rm I}$, i.e.
the logarithmic $\Omega$ and $\DL$ resolution of scenario I. The middle panel
shows $\rm{log_{10}}\, \Delta\Omega^{\rm II}$ and $\rm{log_{10}}\,
\left(\Delta\DL/\DL\right)^{\rm II}$, the logarithmic resolutions of scenario
II. The bottom panel shows $\rm{log_{10}}\, \left(\Delta\Omega^{\rm
I}/\Delta\Omega^{\rm II}\right)$ and $\rm{log_{10}}\, \left(\Delta\DL^{\rm
I}/\Delta\DL^{\rm II}\right)$, the logarithmic ratio of scenario I to scenario
II for both $\Omega$ and $\DL$ resolutions.  \label{fig:OmegaSkymap}}
\end{figure*}

The impact of the sources' location is illustrated in
Figs.~\ref{fig:calMetaSkymap} and~\ref{fig:OmegaSkymap}. We do not show the sky
maps for scenarios III, IV, V, and VI because the results of them are similar to
scenario II.

Figure~\ref{fig:calMetaSkymap} shows the $\calM$ and $\eta$ resolution based on
$(\ell,|m|) = (2,2)$ mode only (denoted with I as indicated in
Table~\ref{tab:HMnotation}) and effect of $(\ell,|m|) = (3,3)$ mode (denoted
with II as indicated in Table~\ref{tab:HMnotation}). The sources' parameters
take the same as Fig.~\ref{fig:SNRskymap}. We find that in scenario I, the
$\calM$ and $\eta$ resolutions are precise, at the order of $10^{-4}$.
Furthermore, in the face-on and face-off areas, the resolution is better than
the edge-on area by factors of $\sim$ 4 and $\sim 8$ for $\calM$ and $\eta$
respectively. The better resolution of the face-on and face-off area is due to
the larger SNR in this area. When $(3,3)$ mode is involved, the $\calM$
resolution can be improved by up to $\sim 1.7$, but in most regions, there is no
improvement. The $\eta$ resolution can be improved by up to $\sim 1.4$. As
expected, the improvement caused by including higher harmonics for $\calM$ and
$\eta$ resolution is marginal. 
To clarify, we have listed the probabilities of precision improvement in Table~\ref{tab:typical_value_calM_eta}, with the source parameters consistent with Fig.~\ref{fig:SNRskymap}. The inclusions of
higher modes do not significantly improve the precisions of these two
parameters, $\calM$ and $\eta$.
Overall, we can not expect a huge improvement in
estimating $\calM$ and $\eta$ by considering higher harmonics.  Nevertheless, if
only considering $(2,2)$ mode, the degeneracy among parameters in the Fisher
matrix is very strong near the face-on and face-off area, but higher harmonics
will break it. Even if we consider the face-on or face-off MBH systems that
theoretically do not contain higher modes (see Fig.~\ref{fig:spin_weighted}),
the inclusion of higher modes can help us better parameter estimations as well.
In other words, the absence of a particular piece of information (mode) is also
information in the data analysis.

\begin{table}
\caption{\label{tab:typical_value_calM_eta} Probability of the typical value of
improvements for chirp mass ($\calM$) and symmetric mass ratio ($\eta$) by including higher harmonics compared with scenario I.}
\begin{ruledtabular}
\begin{tabular}{ccccccc}
Improvement & $1.2$ & $1.4$ & $1.6$ &$1.8$  \\
\hline
$\calM$ & & & &\\
II  & 77.83\% & 12.21\% & 3.12\% & 0\%  \\
III & 86.36\% & 14.71\% & 4.52\% & 0\%  \\
IV  & 86.62\% & 14.81\% & 4.59\% & 0\%  \\
V   & 87.11\% & 15.46\% & 4.79\% & 0\%  \\
VI  & 87.34\% & 15.33\% & 4.85\% & 0\%  \\
\\
\hline
$\eta$& & & &\\
II  & 54.07\% & 0\%     & 0\% & 0\%  \\
III & 67.12\% & 26.33\% & 0\% & 0\%  \\
IV  & 67.51\% & 28.78\% & 0\% & 0\%  \\
V   & 68.23\% & 35.84\% & 0\% & 0\%  \\
VI  & 69.08\% & 33.60\% & 0\% & 0\%  \\
\end{tabular}
\end{ruledtabular}
\end{table}

We also illustrate the angular and luminosity distance resolution of scenario I
and the $(\ell,|m|) = (3,3)$ mode effect in Fig.~\ref{fig:OmegaSkymap}.  The
inclusion of $(\ell,|m|) = (3,3)$ mode results in a dramatic increase in
resolution in most areas except for the near edge-on areas, as which can be
foreseen from Figs.~\ref{fig:total_mass} and~\ref{fig:mass_ratio}. The $\Omega$
and $\DL$ resolutions can be improved by factors $\sim 10^3$ and $\sim 10^5$,
respectively. Even though the inclusion of $(3,3)$ mode can significantly
increase the angular resolution in most areas, the most sensitive resolution of
scenario II is almost the same as scenario I. In other words, $(3,3)$ mode
significantly improves the resolution where scenario I is poor. Nevertheless,
the degeneration between parameters is strong near the face-on and face-off
areas, and we take the results as an optimistic case.

While the $\DL$ resolution improvement can reach a factor of $10^4$, which is
more than $10$ times better than $\Omega$ resolution improvement,
Figure.~\ref{fig:OmegaSkymap} indicates the improvement in $\Omega$ resolution is
more promising. To better address it, we list the probability of the typical
value for precision improvements in Table~\ref{tab:typical value}. The sources'
parameters are consistent with Fig.~\ref{fig:SNRskymap} since it comes from
Fig.~\ref{fig:OmegaSkymap}. Besides the $(3,3)$
mode, the $(4,4)$ mode contributes the most to the precision improvement, and
other higher harmonics contribute only slightly. Half of the $\DL$ resolution
improvements exceed 10, with approximately a 3\%  surpassing over $10^3$. For
the angular resolution, over 70\% of the improvements exceed 10, with around
10\% surpassing over $10^3$. Apart from the $(3,3)$ mode, the other modes also
have a slight effect, particularly when it comes to angular resolution.

\begin{table}
\caption{\label{tab:typical value}Same as
Table~\ref{tab:typical_value_calM_eta}, but for luminosity distance
$\left(\DL\right)$ and angular resolution ($\Omega$).}
\begin{ruledtabular}
\begin{tabular}{ccccccc}
Improvement & $10$ & $10^2$ & $10^3$ &$10^4$  \\
\hline
$\DL$ & & & &\\
II  & 47.62\% & 20.08\% & 3.39\% & 0.26\%  \\
III & 51.13\% & 21.81\% & 3.45\% & 0.26\%  \\
IV  & 51.36\% & 21.48\% & 3.35\% & 0.26\%  \\
V   & 52.70\% & 21.45\% & 3.48\% & 0.26\% \\
VI  & 51.53\% & 21.29\% & 3.48\% & 0.26\%  \\
\\
\hline
$\Omega$& & & &\\
II  & 68.35\% & 46.03\% & 8.56\% & 0\% \\
III & 71.97\% & 50.39\% & 9.67\% & 0\%  \\
IV  & 72.33\% & 50.39\% & 9.05\% & 0\%  \\
V   & 73.50\% & 51.66\% & 10.12\% & 0\% \\
VI  & 72.43\% & 50.81\% & 10.32\% & 0\% \\
\end{tabular}
\end{ruledtabular}
\end{table}

\subsection{The implication for the dark sirens}
\label{subsection:sirens}

In Sec.~\ref{subsection:skymap}, we show the improvement of the angular and
luminosity distance resolution by including higher harmonics. These improvements
can greatly enhance the probability of identifying the host galaxy of MBH.
Meanwhile, Such a dramatic improvement will substantially reduce the error
caused by $\DL$ in constraining $H_0$ and probe cosmology.

The localization of MBH can be mainly searched through the dark siren with the
information of host galaxies. However, usually, the information derived from
parameter estimation is not enough to identify them.

\begin{figure}[htp]
    \includegraphics[scale=0.55]{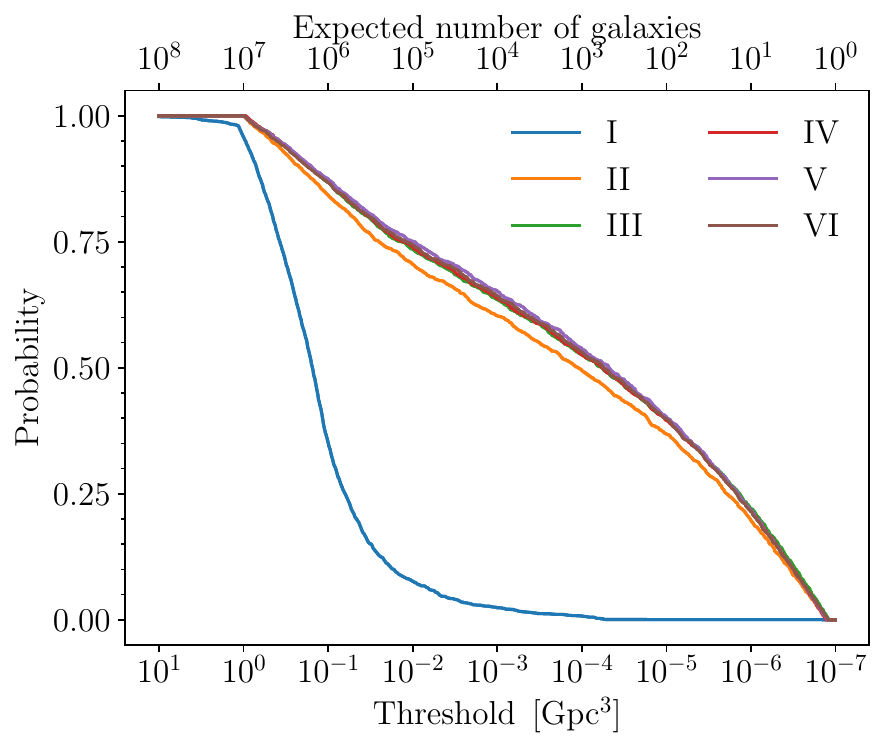}
\caption{\label{fig:probability}The probability of identifying the host galaxy
as a function of the threshold volume for the same parameters as in
Fig.~\ref{fig:SNRskymap}. Each curve corresponds to a different scenario, which
is consistent with  Table~\ref{tab:HMnotation}. The top axis represents the
expected number of host galaxies within the threshold volume, given an assumed
number density of galaxies of $0.01 \,\rm Mpc^{-3}$.}
\end{figure}

To investigate the probability of precise source localization, we define the
volume $\Delta\mathcal{V}$ by
\begin{equation}
    \Delta\mathcal{V} \simeq  \Delta\Omega \frac{\Delta\DL}{\DL}\DL^3 \,.
    \label{eqn:volume}
\end{equation}
Here, $\Delta\mathcal{V}$ denotes the uncertainty volume of the host galaxies'
location. With the assumption that the host galaxies are homogeneous and
isotropic, and the host galaxy can be identified in the threshold volume, then
the probability of identifying the host galaxy can be regarded as a function of
the threshold volume, i.e. the probability that $\Delta\mathcal{V}$ is smaller
than the threshold volume. Note that the average number density of the
Milky-Way-like galaxy is $\sim \rm 0.01\,
Mpc^{-3}$~\cite{Kopparapu:2007ib,LIGOScientific:2010nhs,Chen:2016tys}.
Therefore, the expected number of host galaxies in the threshold volume can be
roughly estimated by $\Delta\mathcal{V}\times 0.01 \rm{Mpc^{-3}}$. The
probability is presented in Fig.~\ref{fig:probability} for the same sources'
parameter as Fig.~\ref{fig:SNRskymap}.

Figure~\ref{fig:probability} shows the inclusion of $(3,3)$ mode will
substantially enhance the probability of identifying the host galaxy, and the
$(4,4)$ mode contributes a visible improvement on this basis, while the rest
modes do not. If the threshold is $\rm 10^{-2}\, Gpc^3$ (the expected number of
host galaxies is $10^5$), considering only $(2,2)$ mode, the probability of
identifying the host galaxy $\sim$ 8\%, while considering higher harmonics it is
over 70\%. When the threshold is $\rm 10^{-4}\, Gpc^3$ (the expected number of
host galaxies is $10^3$), the probability is $\sim$ 7\% if only considering
$(2,2)$ mode, while including higher harmonics it is over 50\%. Even if the
threshold is $\rm 10^{-6}\,  Gpc^3$ (the expected number of host galaxies is
$10$), the probability is around 20\% if higher harmonics are involved. The
$(3,3)$ mode is vital for helping identify the host galaxy, while the $(4,4)$
mode contributes slightly, and the other higher harmonics can be omitted in this
sense. Moreover, the precise location is also helpful for early warning and EM
counterparts' search. Thus, it is necessary to include $(3,3)$ mode at least
into GW data analysis for MBH systems. 

\section{Conclusion}
\label{sec:Conclusion}

BH in the universe is the window to its host and the key to the treasure chest
of cosmology. In this paper, we investigate the impact of various parameters
such as mass, mass ratio, and source location. Moreover, we highlight the
significant effect of higher harmonics on parameter estimation. The $(3,3)$ mode
is the most significant subdominant mode of GWs, the $(4,4)$ mode has a slight
effect on parameter extraction, and the other higher harmonics can be omitted
for LISA.

The main conclusions are summarized in the following,
\begin{enumerate}[(i)]

\item For face-on or face-off MBH systems, in principle, their GW waveforms can
be regarded almost as a $(2,2)$-mode composition (see
Fig.~\ref{fig:spin_weighted}). However, the inclusion of the $(3,3)$ mode is
still necessary for these systems to break the degeneracy in the parameter
estimation. The information about the absence of a particular mode also
contributes to the parameter estimation.

\item Including the $(3,3)$ mode in the data analysis only has a slight effect
on the $\calM$ and $\eta$ estimation. The improvements for those parameters are
more related to the SNRs other than the inclusion of higher modes. Specifically,
including the $(3,3)$ mode leads to the measurements of $\calM$ and $\eta$
improved by a factor $\sim 2$, roughly.

\item Including the $(3,3)$ mode in data analysis significantly affect the $\DL$
and $\Omega$ estimation. Including the $(3,3)$ mode causes improvements in most
of $10^3$-times in angular resolution and $10^4$-times in luminosity distance
resolution. When fixed the redshift $z=1$, relative to $(2,2)$ mode only, the
precisions on luminosity distance and angular resolution from 50\% MBH binary
systems are improved by factors of $10$ and $10^2$, respectively.

\item Including the $(3,3)$ mode will dramatically enhance the probability of
source localization. When the threshold volume is $10^{-2}\, \rm{Gpc^3}$ (the
expected number of host galaxies is $10^5$), including the $(3,3)$ mode will
arise the probability up to 70\%, while less than 8\%  when only considering
$(2,2)$ mode. 

\end{enumerate}

Higher harmonics have a great performance in extracting $\DL$ and $\Omega$. The
great improvement in source localization may constrain $H_0$ precisely to
arbitrate the existing $H_0$ tension problem. GWs emitted from CBCs are proposed
as ``sirens'' to probe $H_0$. To precisely probe the $H_0$, precise $\DL$ and
source localization are vital. $\DL$ inference is directly from GWs, and source
localization is for identifying the host galaxy and obtaining $z$.  However, the
``standard sirens'' are relatively rare in the universe, which will limit the
effect on probing $H_0$.  Besides, the ``dark sirens'' will face the
degeneration between parameters, which typically becomes severe near the face-on
and face-off areas.  This will significantly weaken the effect of probing $H_0$
by the dark sirens. Therefore, the $H_0$ measurement by ``sirens'' is not
precise enough at present, and even the number of detected dark sirens is near 2
orders of magnitude that of standard sirens. For now, the contribution of
constraining $H_0$ is dominated by only one standard siren,
GW170817~\cite{Bulla:2022ppy, LIGOScientific:2021aug}.

Our study focuses on the effect of higher harmonics. We find the effect of
$(3,3)$ mode is the most significant and leads to a significant improvement in
the inferring source localization,  which may identify the unique host galaxy of
the ``dark sirens''.  Thus it may play a pivotal role in black hole physics,
astronomy, and cosmology. Moreover, our analysis can be applied to other similar
space-borne GW detectors, such as Taiji~\cite{TaijiScientific:2021qgx},
TianQin~\cite{TianQin:2015yph}, and DECIGO~\cite{Sato:2017dkf}. In the future,
we expect that the inclusion of eccentricity and precession may enhance breaking
degeneration further and increase the probability of identifying the host
galaxies. Multi-band and multi-detector analysis may also have a visible effect.

\acknowledgments 
We thank the anonymous referee for constructive comments that improved the work.
This work was supported in part by the National Key Research and Development
Program of China Grant (No. 2021YFC2203001) and in part by the National Natural
Science Foundation of China (No.~11920101003, 12021003, 12005016, 12147177,
11975027 and 11991053 ). J.\ Zhao is supported by the ``LiYun'' Postdoctoral
Fellowship of Beijing Normal University.  Z.\ Cao is supported by ``the
Interdiscipline Research Funds of Beijing Normal University" and CAS Project for
Young Scientists in Basic Research YSBR-006. Y.\ Gong is supported by China
Scholarship Council (CSC). L.\ Shao is supported by the National SKA Program of
China (2020SKA0120300) and the Max Planck Partner Group Program funded by the
Max Planck Society.


\begin{thebibliography}{70}%
  \makeatletter
  \providecommand \@ifxundefined [1]{%
   \@ifx{#1\undefined}
  }%
  \providecommand \@ifnum [1]{%
   \ifnum #1\expandafter \@firstoftwo
   \else \expandafter \@secondoftwo
   \fi
  }%
  \providecommand \@ifx [1]{%
   \ifx #1\expandafter \@firstoftwo
   \else \expandafter \@secondoftwo
   \fi
  }%
  \providecommand \natexlab [1]{#1}%
  \providecommand \enquote  [1]{``#1''}%
  \providecommand \bibnamefont  [1]{#1}%
  \providecommand \bibfnamefont [1]{#1}%
  \providecommand \citenamefont [1]{#1}%
  \providecommand \href@noop [0]{\@secondoftwo}%
  \providecommand \href [0]{\begingroup \@sanitize@url \@href}%
  \providecommand \@href[1]{\@@startlink{#1}\@@href}%
  \providecommand \@@href[1]{\endgroup#1\@@endlink}%
  \providecommand \@sanitize@url [0]{\catcode `\\12\catcode `\$12\catcode `\&12\catcode `\#12\catcode `\^12\catcode `\_12\catcode `\%12\relax}%
  \providecommand \@@startlink[1]{}%
  \providecommand \@@endlink[0]{}%
  \providecommand \url  [0]{\begingroup\@sanitize@url \@url }%
  \providecommand \@url [1]{\endgroup\@href {#1}{\urlprefix }}%
  \providecommand \urlprefix  [0]{URL }%
  \providecommand \Eprint [0]{\href }%
  \providecommand \doibase [0]{https://doi.org/}%
  \providecommand \selectlanguage [0]{\@gobble}%
  \providecommand \bibinfo  [0]{\@secondoftwo}%
  \providecommand \bibfield  [0]{\@secondoftwo}%
  \providecommand \translation [1]{[#1]}%
  \providecommand \BibitemOpen [0]{}%
  \providecommand \bibitemStop [0]{}%
  \providecommand \bibitemNoStop [0]{.\EOS\space}%
  \providecommand \EOS [0]{\spacefactor3000\relax}%
  \providecommand \BibitemShut  [1]{\csname bibitem#1\endcsname}%
  \let\auto@bib@innerbib\@empty
  \bibitem [{\citenamefont {Volonteri}(2010)}]{Volonteri:2010wz}%
    \BibitemOpen
    \bibfield  {author} {\bibinfo {author} {\bibfnamefont {M.}~\bibnamefont {Volonteri}},\ }\bibfield  {title} {\bibinfo {title} {{Formation of Supermassive Black Holes}},\ }\href {https://doi.org/10.1007/s00159-010-0029-x} {\bibfield  {journal} {\bibinfo  {journal} {Astron. Astrophys. Rev.}\ }\textbf {\bibinfo {volume} {18}},\ \bibinfo {pages} {279} (\bibinfo {year} {2010})},\ \Eprint {https://arxiv.org/abs/1003.4404} {arXiv:1003.4404 [astro-ph.CO]} \BibitemShut {NoStop}%
  \bibitem [{\citenamefont {Kormendy}\ and\ \citenamefont {Ho}(2013)}]{Kormendy:2013dxa}%
    \BibitemOpen
    \bibfield  {author} {\bibinfo {author} {\bibfnamefont {J.}~\bibnamefont {Kormendy}}\ and\ \bibinfo {author} {\bibfnamefont {L.~C.}\ \bibnamefont {Ho}},\ }\bibfield  {title} {\bibinfo {title} {{Coevolution (Or Not) of Supermassive Black Holes and Host Galaxies}},\ }\href {https://doi.org/10.1146/annurev-astro-082708-101811} {\bibfield  {journal} {\bibinfo  {journal} {Ann. Rev. Astron. Astrophys.}\ }\textbf {\bibinfo {volume} {51}},\ \bibinfo {pages} {511} (\bibinfo {year} {2013})},\ \Eprint {https://arxiv.org/abs/1304.7762} {arXiv:1304.7762 [astro-ph.CO]} \BibitemShut {NoStop}%
  \bibitem [{\citenamefont {Alexander}(2017)}]{Alexander:2017rvg}%
    \BibitemOpen
    \bibfield  {author} {\bibinfo {author} {\bibfnamefont {T.}~\bibnamefont {Alexander}},\ }\bibfield  {title} {\bibinfo {title} {{Stellar Dynamics and Stellar Phenomena Near a Massive Black Hole}},\ }\href {https://doi.org/10.1146/annurev-astro-091916-055306} {\bibfield  {journal} {\bibinfo  {journal} {Ann. Rev. Astron. Astrophys.}\ }\textbf {\bibinfo {volume} {55}},\ \bibinfo {pages} {17} (\bibinfo {year} {2017})},\ \Eprint {https://arxiv.org/abs/1701.04762} {arXiv:1701.04762 [astro-ph.GA]} \BibitemShut {NoStop}%
  \bibitem [{\citenamefont {Barack}\ \emph {et~al.}(2019)\citenamefont {Barack} \emph {et~al.}}]{Barack:2018yly}%
    \BibitemOpen
    \bibfield  {author} {\bibinfo {author} {\bibfnamefont {L.}~\bibnamefont {Barack}} \emph {et~al.},\ }\bibfield  {title} {\bibinfo {title} {{Black holes, gravitational waves and fundamental physics: a roadmap}},\ }\href {https://doi.org/10.1088/1361-6382/ab0587} {\bibfield  {journal} {\bibinfo  {journal} {Class. Quant. Grav.}\ }\textbf {\bibinfo {volume} {36}},\ \bibinfo {pages} {143001} (\bibinfo {year} {2019})},\ \Eprint {https://arxiv.org/abs/1806.05195} {arXiv:1806.05195 [gr-qc]} \BibitemShut {NoStop}%
  \bibitem [{\citenamefont {Schutz}(1986)}]{Schutz:1986gp}%
    \BibitemOpen
    \bibfield  {author} {\bibinfo {author} {\bibfnamefont {B.~F.}\ \bibnamefont {Schutz}},\ }\bibfield  {title} {\bibinfo {title} {{Determining the Hubble Constant from Gravitational Wave Observations}},\ }\href {https://doi.org/10.1038/323310a0} {\bibfield  {journal} {\bibinfo  {journal} {Nature}\ }\textbf {\bibinfo {volume} {323}},\ \bibinfo {pages} {310} (\bibinfo {year} {1986})}\BibitemShut {NoStop}%
  \bibitem [{\citenamefont {Holz}\ and\ \citenamefont {Hughes}(2005)}]{Holz:2005df}%
    \BibitemOpen
    \bibfield  {author} {\bibinfo {author} {\bibfnamefont {D.~E.}\ \bibnamefont {Holz}}\ and\ \bibinfo {author} {\bibfnamefont {S.~A.}\ \bibnamefont {Hughes}},\ }\bibfield  {title} {\bibinfo {title} {{Using gravitational-wave standard sirens}},\ }\href {https://doi.org/10.1086/431341} {\bibfield  {journal} {\bibinfo  {journal} {Astrophys. J.}\ }\textbf {\bibinfo {volume} {629}},\ \bibinfo {pages} {15} (\bibinfo {year} {2005})},\ \Eprint {https://arxiv.org/abs/astro-ph/0504616} {arXiv:astro-ph/0504616} \BibitemShut {NoStop}%
  \bibitem [{\citenamefont {Arun}\ \emph {et~al.}(2022)\citenamefont {Arun} \emph {et~al.}}]{LISA:2022kgy}%
    \BibitemOpen
    \bibfield  {author} {\bibinfo {author} {\bibfnamefont {K.~G.}\ \bibnamefont {Arun}} \emph {et~al.} (\bibinfo {collaboration} {LISA}),\ }\bibfield  {title} {\bibinfo {title} {{New horizons for fundamental physics with LISA}},\ }\href {https://doi.org/10.1007/s41114-022-00036-9} {\bibfield  {journal} {\bibinfo  {journal} {Living Rev. Rel.}\ }\textbf {\bibinfo {volume} {25}},\ \bibinfo {pages} {4} (\bibinfo {year} {2022})},\ \Eprint {https://arxiv.org/abs/2205.01597} {arXiv:2205.01597 [gr-qc]} \BibitemShut {NoStop}%
  \bibitem [{\citenamefont {Bayle}\ \emph {et~al.}(2022)\citenamefont {Bayle}, \citenamefont {Bonga}, \citenamefont {Caprini}, \citenamefont {Doneva}, \citenamefont {Muratore}, \citenamefont {Petiteau}, \citenamefont {Rossi},\ and\ \citenamefont {Shao}}]{Bayle:2022hvs}%
    \BibitemOpen
    \bibfield  {author} {\bibinfo {author} {\bibfnamefont {J.-B.}\ \bibnamefont {Bayle}}, \bibinfo {author} {\bibfnamefont {B.}~\bibnamefont {Bonga}}, \bibinfo {author} {\bibfnamefont {C.}~\bibnamefont {Caprini}}, \bibinfo {author} {\bibfnamefont {D.}~\bibnamefont {Doneva}}, \bibinfo {author} {\bibfnamefont {M.}~\bibnamefont {Muratore}}, \bibinfo {author} {\bibfnamefont {A.}~\bibnamefont {Petiteau}}, \bibinfo {author} {\bibfnamefont {E.}~\bibnamefont {Rossi}},\ and\ \bibinfo {author} {\bibfnamefont {L.}~\bibnamefont {Shao}},\ }\bibfield  {title} {\bibinfo {title} {{Overview and progress on the Laser Interferometer Space Antenna mission}},\ }\href {https://doi.org/10.1038/s41550-022-01847-0} {\bibfield  {journal} {\bibinfo  {journal} {Nature Astron.}\ }\textbf {\bibinfo {volume} {6}},\ \bibinfo {pages} {1334} (\bibinfo {year} {2022})}\BibitemShut {NoStop}%
  \bibitem [{\citenamefont {Amaro-Seoane}\ \emph {et~al.}(2017)\citenamefont {Amaro-Seoane} \emph {et~al.}}]{LISA:2017pwj}%
    \BibitemOpen
    \bibfield  {author} {\bibinfo {author} {\bibfnamefont {P.}~\bibnamefont {Amaro-Seoane}} \emph {et~al.} (\bibinfo {collaboration} {LISA}),\ }\bibfield  {title} {\bibinfo {title} {{Laser Interferometer Space Antenna}},\ }\Eprint {https://arxiv.org/abs/1702.00786} {arXiv:1702.00786 [astro-ph.IM]}  (\bibinfo {year} {2017})\BibitemShut {NoStop}%
  \bibitem [{\citenamefont {Aasi}\ \emph {et~al.}(2015)\citenamefont {Aasi} \emph {et~al.}}]{LIGOScientific:2014pky}%
    \BibitemOpen
    \bibfield  {author} {\bibinfo {author} {\bibfnamefont {J.}~\bibnamefont {Aasi}} \emph {et~al.} (\bibinfo {collaboration} {LIGO Scientific}),\ }\bibfield  {title} {\bibinfo {title} {{Advanced LIGO}},\ }\href {https://doi.org/10.1088/0264-9381/32/7/074001} {\bibfield  {journal} {\bibinfo  {journal} {Class. Quant. Grav.}\ }\textbf {\bibinfo {volume} {32}},\ \bibinfo {pages} {074001} (\bibinfo {year} {2015})},\ \Eprint {https://arxiv.org/abs/1411.4547} {arXiv:1411.4547 [gr-qc]} \BibitemShut {NoStop}%
  \bibitem [{\citenamefont {Acernese}\ \emph {et~al.}(2015)\citenamefont {Acernese} \emph {et~al.}}]{VIRGO:2014yos}%
    \BibitemOpen
    \bibfield  {author} {\bibinfo {author} {\bibfnamefont {F.}~\bibnamefont {Acernese}} \emph {et~al.} (\bibinfo {collaboration} {VIRGO}),\ }\bibfield  {title} {\bibinfo {title} {{Advanced Virgo: a second-generation interferometric gravitational wave detector}},\ }\href {https://doi.org/10.1088/0264-9381/32/2/024001} {\bibfield  {journal} {\bibinfo  {journal} {Class. Quant. Grav.}\ }\textbf {\bibinfo {volume} {32}},\ \bibinfo {pages} {024001} (\bibinfo {year} {2015})},\ \Eprint {https://arxiv.org/abs/1408.3978} {arXiv:1408.3978 [gr-qc]} \BibitemShut {NoStop}%
  \bibitem [{\citenamefont {Akutsu}\ \emph {et~al.}(2021)\citenamefont {Akutsu} \emph {et~al.}}]{KAGRA:2020agh}%
    \BibitemOpen
    \bibfield  {author} {\bibinfo {author} {\bibfnamefont {T.}~\bibnamefont {Akutsu}} \emph {et~al.} (\bibinfo {collaboration} {KAGRA}),\ }\bibfield  {title} {\bibinfo {title} {{Overview of KAGRA: Calibration, detector characterization, physical environmental monitors, and the geophysics interferometer}},\ }\href {https://doi.org/10.1093/ptep/ptab018} {\bibfield  {journal} {\bibinfo  {journal} {PTEP}\ }\textbf {\bibinfo {volume} {2021}},\ \bibinfo {pages} {05A102} (\bibinfo {year} {2021})},\ \Eprint {https://arxiv.org/abs/2009.09305} {arXiv:2009.09305 [gr-qc]} \BibitemShut {NoStop}%
  \bibitem [{\citenamefont {Cutler}(1998)}]{Cutler:1997ta}%
    \BibitemOpen
    \bibfield  {author} {\bibinfo {author} {\bibfnamefont {C.}~\bibnamefont {Cutler}},\ }\bibfield  {title} {\bibinfo {title} {{Angular resolution of the LISA gravitational wave detector}},\ }\href {https://doi.org/10.1103/PhysRevD.57.7089} {\bibfield  {journal} {\bibinfo  {journal} {Phys. Rev. D}\ }\textbf {\bibinfo {volume} {57}},\ \bibinfo {pages} {7089} (\bibinfo {year} {1998})},\ \Eprint {https://arxiv.org/abs/gr-qc/9703068} {arXiv:gr-qc/9703068} \BibitemShut {NoStop}%
  \bibitem [{\citenamefont {Abbott}\ \emph {et~al.}(2019)\citenamefont {Abbott} \emph {et~al.}}]{LIGOScientific:2018mvr}%
    \BibitemOpen
    \bibfield  {author} {\bibinfo {author} {\bibfnamefont {B.~P.}\ \bibnamefont {Abbott}} \emph {et~al.} (\bibinfo {collaboration} {LIGO Scientific, Virgo}),\ }\bibfield  {title} {\bibinfo {title} {{GWTC-1: A Gravitational-Wave Transient Catalog of Compact Binary Mergers Observed by LIGO and Virgo during the First and Second Observing Runs}},\ }\href {https://doi.org/10.1103/PhysRevX.9.031040} {\bibfield  {journal} {\bibinfo  {journal} {Phys. Rev. X}\ }\textbf {\bibinfo {volume} {9}},\ \bibinfo {pages} {031040} (\bibinfo {year} {2019})},\ \Eprint {https://arxiv.org/abs/1811.12907} {arXiv:1811.12907 [astro-ph.HE]} \BibitemShut {NoStop}%
  \bibitem [{\citenamefont {Abbott}\ \emph {et~al.}(2021{\natexlab{a}})\citenamefont {Abbott} \emph {et~al.}}]{LIGOScientific:2020ibl}%
    \BibitemOpen
    \bibfield  {author} {\bibinfo {author} {\bibfnamefont {R.}~\bibnamefont {Abbott}} \emph {et~al.} (\bibinfo {collaboration} {LIGO Scientific, Virgo}),\ }\bibfield  {title} {\bibinfo {title} {{GWTC-2: Compact Binary Coalescences Observed by LIGO and Virgo During the First Half of the Third Observing Run}},\ }\href {https://doi.org/10.1103/PhysRevX.11.021053} {\bibfield  {journal} {\bibinfo  {journal} {Phys. Rev. X}\ }\textbf {\bibinfo {volume} {11}},\ \bibinfo {pages} {021053} (\bibinfo {year} {2021}{\natexlab{a}})},\ \Eprint {https://arxiv.org/abs/2010.14527} {arXiv:2010.14527 [gr-qc]} \BibitemShut {NoStop}%
  \bibitem [{\citenamefont {Abbott}\ \emph {et~al.}(2021{\natexlab{b}})\citenamefont {Abbott} \emph {et~al.}}]{LIGOScientific:2021usb}%
    \BibitemOpen
    \bibfield  {author} {\bibinfo {author} {\bibfnamefont {R.}~\bibnamefont {Abbott}} \emph {et~al.} (\bibinfo {collaboration} {LIGO Scientific, VIRGO}),\ }\bibfield  {title} {\bibinfo {title} {{GWTC-2.1: Deep Extended Catalog of Compact Binary Coalescences Observed by LIGO and Virgo During the First Half of the Third Observing Run}},\ }\Eprint {https://arxiv.org/abs/2108.01045} {arXiv:2108.01045 [gr-qc]}  (\bibinfo {year} {2021}{\natexlab{b}})\BibitemShut {NoStop}%
  \bibitem [{\citenamefont {Abbott}\ \emph {et~al.}(2021{\natexlab{c}})\citenamefont {Abbott} \emph {et~al.}}]{LIGOScientific:2021djp}%
    \BibitemOpen
    \bibfield  {author} {\bibinfo {author} {\bibfnamefont {R.}~\bibnamefont {Abbott}} \emph {et~al.} (\bibinfo {collaboration} {LIGO Scientific, VIRGO, KAGRA}),\ }\bibfield  {title} {\bibinfo {title} {{GWTC-3: Compact Binary Coalescences Observed by LIGO and Virgo During the Second Part of the Third Observing Run}},\ }\Eprint {https://arxiv.org/abs/2111.03606} {arXiv:2111.03606 [gr-qc]}  (\bibinfo {year} {2021}{\natexlab{c}})\BibitemShut {NoStop}%
  \bibitem [{\citenamefont {Berti}\ \emph {et~al.}(2005)\citenamefont {Berti}, \citenamefont {Buonanno},\ and\ \citenamefont {Will}}]{Berti:2004bd}%
    \BibitemOpen
    \bibfield  {author} {\bibinfo {author} {\bibfnamefont {E.}~\bibnamefont {Berti}}, \bibinfo {author} {\bibfnamefont {A.}~\bibnamefont {Buonanno}},\ and\ \bibinfo {author} {\bibfnamefont {C.~M.}\ \bibnamefont {Will}},\ }\bibfield  {title} {\bibinfo {title} {{Estimating spinning binary parameters and testing alternative theories of gravity with LISA}},\ }\href {https://doi.org/10.1103/PhysRevD.71.084025} {\bibfield  {journal} {\bibinfo  {journal} {Phys. Rev. D}\ }\textbf {\bibinfo {volume} {71}},\ \bibinfo {pages} {084025} (\bibinfo {year} {2005})},\ \Eprint {https://arxiv.org/abs/gr-qc/0411129} {arXiv:gr-qc/0411129} \BibitemShut {NoStop}%
  \bibitem [{\citenamefont {Thorne}(1980)}]{Thorne:1980ru}%
    \BibitemOpen
    \bibfield  {author} {\bibinfo {author} {\bibfnamefont {K.~S.}\ \bibnamefont {Thorne}},\ }\bibfield  {title} {\bibinfo {title} {{Multipole Expansions of Gravitational Radiation}},\ }\href {https://doi.org/10.1103/RevModPhys.52.299} {\bibfield  {journal} {\bibinfo  {journal} {Rev. Mod. Phys.}\ }\textbf {\bibinfo {volume} {52}},\ \bibinfo {pages} {299} (\bibinfo {year} {1980})}\BibitemShut {NoStop}%
  \bibitem [{\citenamefont {Blanchet}(2014)}]{Blanchet:2013haa}%
    \BibitemOpen
    \bibfield  {author} {\bibinfo {author} {\bibfnamefont {L.}~\bibnamefont {Blanchet}},\ }\bibfield  {title} {\bibinfo {title} {{Gravitational Radiation from Post-Newtonian Sources and Inspiralling Compact Binaries}},\ }\href {https://doi.org/10.12942/lrr-2014-2} {\bibfield  {journal} {\bibinfo  {journal} {Living Rev. Rel.}\ }\textbf {\bibinfo {volume} {17}},\ \bibinfo {pages} {2} (\bibinfo {year} {2014})},\ \Eprint {https://arxiv.org/abs/1310.1528} {arXiv:1310.1528 [gr-qc]} \BibitemShut {NoStop}%
  \bibitem [{\citenamefont {London}\ \emph {et~al.}(2018)\citenamefont {London}, \citenamefont {Khan}, \citenamefont {Fauchon-Jones}, \citenamefont {Garc\'\i{}a}, \citenamefont {Hannam}, \citenamefont {Husa}, \citenamefont {Jim\'enez-Forteza}, \citenamefont {Kalaghatgi}, \citenamefont {Ohme},\ and\ \citenamefont {Pannarale}}]{London:2017bcn}%
    \BibitemOpen
    \bibfield  {author} {\bibinfo {author} {\bibfnamefont {L.}~\bibnamefont {London}}, \bibinfo {author} {\bibfnamefont {S.}~\bibnamefont {Khan}}, \bibinfo {author} {\bibfnamefont {E.}~\bibnamefont {Fauchon-Jones}}, \bibinfo {author} {\bibfnamefont {C.}~\bibnamefont {Garc\'\i{}a}}, \bibinfo {author} {\bibfnamefont {M.}~\bibnamefont {Hannam}}, \bibinfo {author} {\bibfnamefont {S.}~\bibnamefont {Husa}}, \bibinfo {author} {\bibfnamefont {X.}~\bibnamefont {Jim\'enez-Forteza}}, \bibinfo {author} {\bibfnamefont {C.}~\bibnamefont {Kalaghatgi}}, \bibinfo {author} {\bibfnamefont {F.}~\bibnamefont {Ohme}},\ and\ \bibinfo {author} {\bibfnamefont {F.}~\bibnamefont {Pannarale}},\ }\bibfield  {title} {\bibinfo {title} {{First higher-multipole model of gravitational waves from spinning and coalescing black-hole binaries}},\ }\href {https://doi.org/10.1103/PhysRevLett.120.161102} {\bibfield  {journal} {\bibinfo  {journal} {Phys. Rev. Lett.}\ }\textbf {\bibinfo {volume} {120}},\ \bibinfo {pages} {161102} (\bibinfo {year} {2018})},\ \Eprint {https://arxiv.org/abs/1708.00404} {arXiv:1708.00404 [gr-qc]} \BibitemShut {NoStop}%
  \bibitem [{\citenamefont {Abbott}\ \emph {et~al.}(2020{\natexlab{a}})\citenamefont {Abbott} \emph {et~al.}}]{LIGOScientific:2020stg}%
    \BibitemOpen
    \bibfield  {author} {\bibinfo {author} {\bibfnamefont {R.}~\bibnamefont {Abbott}} \emph {et~al.} (\bibinfo {collaboration} {LIGO Scientific, Virgo}),\ }\bibfield  {title} {\bibinfo {title} {{GW190412: Observation of a Binary-Black-Hole Coalescence with Asymmetric Masses}},\ }\href {https://doi.org/10.1103/PhysRevD.102.043015} {\bibfield  {journal} {\bibinfo  {journal} {Phys. Rev. D}\ }\textbf {\bibinfo {volume} {102}},\ \bibinfo {pages} {043015} (\bibinfo {year} {2020}{\natexlab{a}})},\ \Eprint {https://arxiv.org/abs/2004.08342} {arXiv:2004.08342 [astro-ph.HE]} \BibitemShut {NoStop}%
  \bibitem [{\citenamefont {Abbott}\ \emph {et~al.}(2020{\natexlab{b}})\citenamefont {Abbott} \emph {et~al.}}]{LIGOScientific:2020zkf}%
    \BibitemOpen
    \bibfield  {author} {\bibinfo {author} {\bibfnamefont {R.}~\bibnamefont {Abbott}} \emph {et~al.} (\bibinfo {collaboration} {LIGO Scientific, Virgo}),\ }\bibfield  {title} {\bibinfo {title} {{GW190814: Gravitational Waves from the Coalescence of a 23 Solar Mass Black Hole with a 2.6 Solar Mass Compact Object}},\ }\href {https://doi.org/10.3847/2041-8213/ab960f} {\bibfield  {journal} {\bibinfo  {journal} {Astrophys. J. Lett.}\ }\textbf {\bibinfo {volume} {896}},\ \bibinfo {pages} {L44} (\bibinfo {year} {2020}{\natexlab{b}})},\ \Eprint {https://arxiv.org/abs/2006.12611} {arXiv:2006.12611 [astro-ph.HE]} \BibitemShut {NoStop}%
  \bibitem [{\citenamefont {Chatziioannou}\ \emph {et~al.}(2019)\citenamefont {Chatziioannou} \emph {et~al.}}]{Chatziioannou:2019dsz}%
    \BibitemOpen
    \bibfield  {author} {\bibinfo {author} {\bibfnamefont {K.}~\bibnamefont {Chatziioannou}} \emph {et~al.},\ }\bibfield  {title} {\bibinfo {title} {{On the properties of the massive binary black hole merger GW170729}},\ }\href {https://doi.org/10.1103/PhysRevD.100.104015} {\bibfield  {journal} {\bibinfo  {journal} {Phys. Rev. D}\ }\textbf {\bibinfo {volume} {100}},\ \bibinfo {pages} {104015} (\bibinfo {year} {2019})},\ \Eprint {https://arxiv.org/abs/1903.06742} {arXiv:1903.06742 [gr-qc]} \BibitemShut {NoStop}%
  \bibitem [{\citenamefont {Krishnendu}\ and\ \citenamefont {Ohme}(2022)}]{Krishnendu:2021cyi}%
    \BibitemOpen
    \bibfield  {author} {\bibinfo {author} {\bibfnamefont {N.~V.}\ \bibnamefont {Krishnendu}}\ and\ \bibinfo {author} {\bibfnamefont {F.}~\bibnamefont {Ohme}},\ }\bibfield  {title} {\bibinfo {title} {{Interplay of spin-precession and higher harmonics in the parameter estimation of binary black holes}},\ }\href {https://doi.org/10.1103/PhysRevD.105.064012} {\bibfield  {journal} {\bibinfo  {journal} {Phys. Rev. D}\ }\textbf {\bibinfo {volume} {105}},\ \bibinfo {pages} {064012} (\bibinfo {year} {2022})},\ \Eprint {https://arxiv.org/abs/2110.00766} {arXiv:2110.00766 [gr-qc]} \BibitemShut {NoStop}%
  \bibitem [{\citenamefont {Wang}\ and\ \citenamefont {Hu}(2022)}]{Wang:2022apn}%
    \BibitemOpen
    \bibfield  {author} {\bibinfo {author} {\bibfnamefont {R.}~\bibnamefont {Wang}}\ and\ \bibinfo {author} {\bibfnamefont {B.}~\bibnamefont {Hu}},\ }\bibfield  {title} {\bibinfo {title} {{LitePIG: A Lite Parameter Inference system for the Gravitational wave in the millihertz band}},\ }\Eprint {https://arxiv.org/abs/2208.13351} {arXiv:2208.13351 [astro-ph.IM]}  (\bibinfo {year} {2022})\BibitemShut {NoStop}%
  \bibitem [{\citenamefont {Gao}\ \emph {et~al.}(2022)\citenamefont {Gao}, \citenamefont {You}, \citenamefont {Gong}, \citenamefont {Zhang},\ and\ \citenamefont {Zhang}}]{Gao:2022hsn}%
    \BibitemOpen
    \bibfield  {author} {\bibinfo {author} {\bibfnamefont {Q.}~\bibnamefont {Gao}}, \bibinfo {author} {\bibfnamefont {Y.}~\bibnamefont {You}}, \bibinfo {author} {\bibfnamefont {Y.}~\bibnamefont {Gong}}, \bibinfo {author} {\bibfnamefont {C.}~\bibnamefont {Zhang}},\ and\ \bibinfo {author} {\bibfnamefont {C.}~\bibnamefont {Zhang}},\ }\bibfield  {title} {\bibinfo {title} {{Testing alternative theories of gravity with space-based gravitational wave detectors}},\ }\Eprint {https://arxiv.org/abs/2212.03789} {arXiv:2212.03789 [gr-qc]}  (\bibinfo {year} {2022})\BibitemShut {NoStop}%
  \bibitem [{\citenamefont {Varma}\ and\ \citenamefont {Ajith}(2017)}]{Varma:2016dnf}%
    \BibitemOpen
    \bibfield  {author} {\bibinfo {author} {\bibfnamefont {V.}~\bibnamefont {Varma}}\ and\ \bibinfo {author} {\bibfnamefont {P.}~\bibnamefont {Ajith}},\ }\bibfield  {title} {\bibinfo {title} {{Effects of nonquadrupole modes in the detection and parameter estimation of black hole binaries with nonprecessing spins}},\ }\href {https://doi.org/10.1103/PhysRevD.96.124024} {\bibfield  {journal} {\bibinfo  {journal} {Phys. Rev. D}\ }\textbf {\bibinfo {volume} {96}},\ \bibinfo {pages} {124024} (\bibinfo {year} {2017})},\ \Eprint {https://arxiv.org/abs/1612.05608} {arXiv:1612.05608 [gr-qc]} \BibitemShut {NoStop}%
  \bibitem [{\citenamefont {Varma}\ \emph {et~al.}(2019)\citenamefont {Varma}, \citenamefont {Field}, \citenamefont {Scheel}, \citenamefont {Blackman}, \citenamefont {Kidder},\ and\ \citenamefont {Pfeiffer}}]{Varma:2018mmi}%
    \BibitemOpen
    \bibfield  {author} {\bibinfo {author} {\bibfnamefont {V.}~\bibnamefont {Varma}}, \bibinfo {author} {\bibfnamefont {S.~E.}\ \bibnamefont {Field}}, \bibinfo {author} {\bibfnamefont {M.~A.}\ \bibnamefont {Scheel}}, \bibinfo {author} {\bibfnamefont {J.}~\bibnamefont {Blackman}}, \bibinfo {author} {\bibfnamefont {L.~E.}\ \bibnamefont {Kidder}},\ and\ \bibinfo {author} {\bibfnamefont {H.~P.}\ \bibnamefont {Pfeiffer}},\ }\bibfield  {title} {\bibinfo {title} {{Surrogate model of hybridized numerical relativity binary black hole waveforms}},\ }\href {https://doi.org/10.1103/PhysRevD.99.064045} {\bibfield  {journal} {\bibinfo  {journal} {Phys. Rev. D}\ }\textbf {\bibinfo {volume} {99}},\ \bibinfo {pages} {064045} (\bibinfo {year} {2019})},\ \Eprint {https://arxiv.org/abs/1812.07865} {arXiv:1812.07865 [gr-qc]} \BibitemShut {NoStop}%
  \bibitem [{\citenamefont {Porter}\ and\ \citenamefont {Cornish}(2008)}]{Porter:2008kn}%
    \BibitemOpen
    \bibfield  {author} {\bibinfo {author} {\bibfnamefont {E.~K.}\ \bibnamefont {Porter}}\ and\ \bibinfo {author} {\bibfnamefont {N.~J.}\ \bibnamefont {Cornish}},\ }\bibfield  {title} {\bibinfo {title} {{The Effect of Higher Harmonic Corrections on the Detection of massive black hole binaries with LISA}},\ }\href {https://doi.org/10.1103/PhysRevD.78.064005} {\bibfield  {journal} {\bibinfo  {journal} {Phys. Rev. D}\ }\textbf {\bibinfo {volume} {78}},\ \bibinfo {pages} {064005} (\bibinfo {year} {2008})},\ \Eprint {https://arxiv.org/abs/0804.0332} {arXiv:0804.0332 [gr-qc]} \BibitemShut {NoStop}%
  \bibitem [{\citenamefont {Trias}\ and\ \citenamefont {Sintes}(2008)}]{Trias:2007fp}%
    \BibitemOpen
    \bibfield  {author} {\bibinfo {author} {\bibfnamefont {M.}~\bibnamefont {Trias}}\ and\ \bibinfo {author} {\bibfnamefont {A.~M.}\ \bibnamefont {Sintes}},\ }\bibfield  {title} {\bibinfo {title} {{LISA observations of supermassive black holes: Parameter estimation using full post-Newtonian inspiral waveforms}},\ }\href {https://doi.org/10.1103/PhysRevD.77.024030} {\bibfield  {journal} {\bibinfo  {journal} {Phys. Rev. D}\ }\textbf {\bibinfo {volume} {77}},\ \bibinfo {pages} {024030} (\bibinfo {year} {2008})},\ \Eprint {https://arxiv.org/abs/0707.4434} {arXiv:0707.4434 [gr-qc]} \BibitemShut {NoStop}%
  \bibitem [{\citenamefont {Arun}\ \emph {et~al.}(2007)\citenamefont {Arun}, \citenamefont {Iyer}, \citenamefont {Sathyaprakash}, \citenamefont {Sinha},\ and\ \citenamefont {Van Den~Broeck}}]{Arun:2007hu}%
    \BibitemOpen
    \bibfield  {author} {\bibinfo {author} {\bibfnamefont {K.~G.}\ \bibnamefont {Arun}}, \bibinfo {author} {\bibfnamefont {B.~R.}\ \bibnamefont {Iyer}}, \bibinfo {author} {\bibfnamefont {B.~S.}\ \bibnamefont {Sathyaprakash}}, \bibinfo {author} {\bibfnamefont {S.}~\bibnamefont {Sinha}},\ and\ \bibinfo {author} {\bibfnamefont {C.}~\bibnamefont {Van Den~Broeck}},\ }\bibfield  {title} {\bibinfo {title} {{Higher signal harmonics, LISA's angular resolution and dark energy}},\ }\href {https://doi.org/10.1103/PhysRevD.76.104016} {\bibfield  {journal} {\bibinfo  {journal} {Phys. Rev. D}\ }\textbf {\bibinfo {volume} {76}},\ \bibinfo {pages} {104016} (\bibinfo {year} {2007})},\ \bibinfo {note} {[Erratum: Phys.Rev.D 76, 129903 (2007)]},\ \Eprint {https://arxiv.org/abs/0707.3920} {arXiv:0707.3920 [astro-ph]} \BibitemShut {NoStop}%
  \bibitem [{\citenamefont {Baibhav}\ \emph {et~al.}(2020)\citenamefont {Baibhav}, \citenamefont {Berti},\ and\ \citenamefont {Cardoso}}]{Baibhav:2020tma}%
    \BibitemOpen
    \bibfield  {author} {\bibinfo {author} {\bibfnamefont {V.}~\bibnamefont {Baibhav}}, \bibinfo {author} {\bibfnamefont {E.}~\bibnamefont {Berti}},\ and\ \bibinfo {author} {\bibfnamefont {V.}~\bibnamefont {Cardoso}},\ }\bibfield  {title} {\bibinfo {title} {{LISA parameter estimation and source localization with higher harmonics of the ringdown}},\ }\href {https://doi.org/10.1103/PhysRevD.101.084053} {\bibfield  {journal} {\bibinfo  {journal} {Phys. Rev. D}\ }\textbf {\bibinfo {volume} {101}},\ \bibinfo {pages} {084053} (\bibinfo {year} {2020})},\ \Eprint {https://arxiv.org/abs/2001.10011} {arXiv:2001.10011 [gr-qc]} \BibitemShut {NoStop}%
  \bibitem [{\citenamefont {Marsat}\ \emph {et~al.}(2021)\citenamefont {Marsat}, \citenamefont {Baker},\ and\ \citenamefont {Dal~Canton}}]{Marsat:2020rtl}%
    \BibitemOpen
    \bibfield  {author} {\bibinfo {author} {\bibfnamefont {S.}~\bibnamefont {Marsat}}, \bibinfo {author} {\bibfnamefont {J.~G.}\ \bibnamefont {Baker}},\ and\ \bibinfo {author} {\bibfnamefont {T.}~\bibnamefont {Dal~Canton}},\ }\bibfield  {title} {\bibinfo {title} {{Exploring the Bayesian parameter estimation of binary black holes with LISA}},\ }\href {https://doi.org/10.1103/PhysRevD.103.083011} {\bibfield  {journal} {\bibinfo  {journal} {Phys. Rev. D}\ }\textbf {\bibinfo {volume} {103}},\ \bibinfo {pages} {083011} (\bibinfo {year} {2021})},\ \Eprint {https://arxiv.org/abs/2003.00357} {arXiv:2003.00357 [gr-qc]} \BibitemShut {NoStop}%
  \bibitem [{\citenamefont {Pratten}\ \emph {et~al.}(2022)\citenamefont {Pratten}, \citenamefont {Klein}, \citenamefont {Moore}, \citenamefont {Middleton}, \citenamefont {Steinle}, \citenamefont {Schmidt},\ and\ \citenamefont {Vecchio}}]{Pratten:2022kug}%
    \BibitemOpen
    \bibfield  {author} {\bibinfo {author} {\bibfnamefont {G.}~\bibnamefont {Pratten}}, \bibinfo {author} {\bibfnamefont {A.}~\bibnamefont {Klein}}, \bibinfo {author} {\bibfnamefont {C.~J.}\ \bibnamefont {Moore}}, \bibinfo {author} {\bibfnamefont {H.}~\bibnamefont {Middleton}}, \bibinfo {author} {\bibfnamefont {N.}~\bibnamefont {Steinle}}, \bibinfo {author} {\bibfnamefont {P.}~\bibnamefont {Schmidt}},\ and\ \bibinfo {author} {\bibfnamefont {A.}~\bibnamefont {Vecchio}},\ }\bibfield  {title} {\bibinfo {title} {{On the LISA science performance in observations of short-lived signals from massive black hole binary coalescences}},\ }\Eprint {https://arxiv.org/abs/2212.02572} {arXiv:2212.02572 [gr-qc]}  (\bibinfo {year} {2022})\BibitemShut {NoStop}%
  \bibitem [{\citenamefont {Katz}(2022)}]{Katz:2021uax}%
    \BibitemOpen
    \bibfield  {author} {\bibinfo {author} {\bibfnamefont {M.~L.}\ \bibnamefont {Katz}},\ }\bibfield  {title} {\bibinfo {title} {{Fully automated end-to-end pipeline for massive black hole binary signal extraction from LISA data}},\ }\href {https://doi.org/10.1103/PhysRevD.105.044055} {\bibfield  {journal} {\bibinfo  {journal} {Phys. Rev. D}\ }\textbf {\bibinfo {volume} {105}},\ \bibinfo {pages} {044055} (\bibinfo {year} {2022})},\ \Eprint {https://arxiv.org/abs/2111.01064} {arXiv:2111.01064 [gr-qc]} \BibitemShut {NoStop}%
  \bibitem [{\citenamefont {Ng}\ \emph {et~al.}(2023)\citenamefont {Ng} \emph {et~al.}}]{Ng:2022vbz}%
    \BibitemOpen
    \bibfield  {author} {\bibinfo {author} {\bibfnamefont {K.~K.~Y.}\ \bibnamefont {Ng}} \emph {et~al.},\ }\bibfield  {title} {\bibinfo {title} {{Measuring properties of primordial black hole mergers at cosmological distances: Effect of higher order modes in gravitational waves}},\ }\href {https://doi.org/10.1103/PhysRevD.107.024041} {\bibfield  {journal} {\bibinfo  {journal} {Phys. Rev. D}\ }\textbf {\bibinfo {volume} {107}},\ \bibinfo {pages} {024041} (\bibinfo {year} {2023})},\ \Eprint {https://arxiv.org/abs/2210.03132} {arXiv:2210.03132 [astro-ph.CO]} \BibitemShut {NoStop}%
  \bibitem [{\citenamefont {Iacovelli}\ \emph {et~al.}(2022)\citenamefont {Iacovelli}, \citenamefont {Mancarella}, \citenamefont {Foffa},\ and\ \citenamefont {Maggiore}}]{Iacovelli:2022mbg}%
    \BibitemOpen
    \bibfield  {author} {\bibinfo {author} {\bibfnamefont {F.}~\bibnamefont {Iacovelli}}, \bibinfo {author} {\bibfnamefont {M.}~\bibnamefont {Mancarella}}, \bibinfo {author} {\bibfnamefont {S.}~\bibnamefont {Foffa}},\ and\ \bibinfo {author} {\bibfnamefont {M.}~\bibnamefont {Maggiore}},\ }\bibfield  {title} {\bibinfo {title} {{GWFAST: A Fisher Information Matrix Python Code for Third-generation Gravitational-wave Detectors}},\ }\href {https://doi.org/10.3847/1538-4365/ac9129} {\bibfield  {journal} {\bibinfo  {journal} {Astrophys. J. Supp.}\ }\textbf {\bibinfo {volume} {263}},\ \bibinfo {pages} {2} (\bibinfo {year} {2022})},\ \Eprint {https://arxiv.org/abs/2207.06910} {arXiv:2207.06910 [astro-ph.IM]} \BibitemShut {NoStop}%
  \bibitem [{\citenamefont {Abbott}\ \emph {et~al.}(2017{\natexlab{a}})\citenamefont {Abbott} \emph {et~al.}}]{LIGOScientific:2017vwq}%
    \BibitemOpen
    \bibfield  {author} {\bibinfo {author} {\bibfnamefont {B.~P.}\ \bibnamefont {Abbott}} \emph {et~al.} (\bibinfo {collaboration} {LIGO Scientific, Virgo}),\ }\bibfield  {title} {\bibinfo {title} {{GW170817: Observation of Gravitational Waves from a Binary Neutron Star Inspiral}},\ }\href {https://doi.org/10.1103/PhysRevLett.119.161101} {\bibfield  {journal} {\bibinfo  {journal} {Phys. Rev. Lett.}\ }\textbf {\bibinfo {volume} {119}},\ \bibinfo {pages} {161101} (\bibinfo {year} {2017}{\natexlab{a}})},\ \Eprint {https://arxiv.org/abs/1710.05832} {arXiv:1710.05832 [gr-qc]} \BibitemShut {NoStop}%
  \bibitem [{\citenamefont {Coulter}\ \emph {et~al.}(2017)\citenamefont {Coulter} \emph {et~al.}}]{Coulter:2017wya}%
    \BibitemOpen
    \bibfield  {author} {\bibinfo {author} {\bibfnamefont {D.~A.}\ \bibnamefont {Coulter}} \emph {et~al.},\ }\bibfield  {title} {\bibinfo {title} {{Swope Supernova Survey 2017a (SSS17a), the Optical Counterpart to a Gravitational Wave Source}},\ }\href {https://doi.org/10.1126/science.aap9811} {\bibfield  {journal} {\bibinfo  {journal} {Science}\ }\textbf {\bibinfo {volume} {358}},\ \bibinfo {pages} {1556} (\bibinfo {year} {2017})},\ \Eprint {https://arxiv.org/abs/1710.05452} {arXiv:1710.05452 [astro-ph.HE]} \BibitemShut {NoStop}%
  \bibitem [{\citenamefont {Abbott}\ \emph {et~al.}(2017{\natexlab{b}})\citenamefont {Abbott} \emph {et~al.}}]{LIGOScientific:2017zic}%
    \BibitemOpen
    \bibfield  {author} {\bibinfo {author} {\bibfnamefont {B.~P.}\ \bibnamefont {Abbott}} \emph {et~al.} (\bibinfo {collaboration} {LIGO Scientific, Virgo, Fermi-GBM, INTEGRAL}),\ }\bibfield  {title} {\bibinfo {title} {{Gravitational Waves and Gamma-rays from a Binary Neutron Star Merger: GW170817 and GRB 170817A}},\ }\href {https://doi.org/10.3847/2041-8213/aa920c} {\bibfield  {journal} {\bibinfo  {journal} {Astrophys. J. Lett.}\ }\textbf {\bibinfo {volume} {848}},\ \bibinfo {pages} {L13} (\bibinfo {year} {2017}{\natexlab{b}})},\ \Eprint {https://arxiv.org/abs/1710.05834} {arXiv:1710.05834 [astro-ph.HE]} \BibitemShut {NoStop}%
  \bibitem [{\citenamefont {Abbott}\ \emph {et~al.}(2017{\natexlab{c}})\citenamefont {Abbott} \emph {et~al.}}]{LIGOScientific:2017ync}%
    \BibitemOpen
    \bibfield  {author} {\bibinfo {author} {\bibfnamefont {B.~P.}\ \bibnamefont {Abbott}} \emph {et~al.} (\bibinfo {collaboration} {LIGO Scientific, Virgo, Fermi GBM, INTEGRAL, IceCube, AstroSat Cadmium Zinc Telluride Imager Team, IPN, Insight-Hxmt, ANTARES, Swift, AGILE Team, 1M2H Team, Dark Energy Camera GW-EM, DES, DLT40, GRAWITA, Fermi-LAT, ATCA, ASKAP, Las Cumbres Observatory Group, OzGrav, DWF (Deeper Wider Faster Program), AST3, CAASTRO, VINROUGE, MASTER, J-GEM, GROWTH, JAGWAR, CaltechNRAO, TTU-NRAO, NuSTAR, Pan-STARRS, MAXI Team, TZAC Consortium, KU, Nordic Optical Telescope, ePESSTO, GROND, Texas Tech University, SALT Group, TOROS, BOOTES, MWA, CALET, IKI-GW Follow-up, H.E.S.S., LOFAR, LWA, HAWC, Pierre Auger, ALMA, Euro VLBI Team, Pi of Sky, Chandra Team at McGill University, DFN, ATLAS Telescopes, High Time Resolution Universe Survey, RIMAS, RATIR, SKA South Africa/MeerKAT}),\ }\bibfield  {title} {\bibinfo {title} {{Multi-messenger Observations of a Binary Neutron Star Merger}},\ }\href {https://doi.org/10.3847/2041-8213/aa91c9} {\bibfield  {journal} {\bibinfo  {journal} {Astrophys. J. Lett.}\ }\textbf {\bibinfo {volume} {848}},\ \bibinfo {pages} {L12} (\bibinfo {year} {2017}{\natexlab{c}})},\ \Eprint {https://arxiv.org/abs/1710.05833} {arXiv:1710.05833 [astro-ph.HE]} \BibitemShut {NoStop}%
  \bibitem [{\citenamefont {Abbott}\ \emph {et~al.}(2017{\natexlab{d}})\citenamefont {Abbott} \emph {et~al.}}]{LIGOScientific:2017adf}%
    \BibitemOpen
    \bibfield  {author} {\bibinfo {author} {\bibfnamefont {B.~P.}\ \bibnamefont {Abbott}} \emph {et~al.} (\bibinfo {collaboration} {LIGO Scientific, Virgo, 1M2H, Dark Energy Camera GW-E, DES, DLT40, Las Cumbres Observatory, VINROUGE, MASTER}),\ }\bibfield  {title} {\bibinfo {title} {{A gravitational-wave standard siren measurement of the Hubble constant}},\ }\href {https://doi.org/10.1038/nature24471} {\bibfield  {journal} {\bibinfo  {journal} {Nature}\ }\textbf {\bibinfo {volume} {551}},\ \bibinfo {pages} {85} (\bibinfo {year} {2017}{\natexlab{d}})},\ \Eprint {https://arxiv.org/abs/1710.05835} {arXiv:1710.05835 [astro-ph.CO]} \BibitemShut {NoStop}%
  \bibitem [{\citenamefont {Aghanim}\ \emph {et~al.}(2020)\citenamefont {Aghanim} \emph {et~al.}}]{Planck:2018vyg}%
    \BibitemOpen
    \bibfield  {author} {\bibinfo {author} {\bibfnamefont {N.}~\bibnamefont {Aghanim}} \emph {et~al.} (\bibinfo {collaboration} {Planck}),\ }\bibfield  {title} {\bibinfo {title} {{Planck 2018 results. VI. Cosmological parameters}},\ }\href {https://doi.org/10.1051/0004-6361/201833910} {\bibfield  {journal} {\bibinfo  {journal} {Astron. Astrophys.}\ }\textbf {\bibinfo {volume} {641}},\ \bibinfo {pages} {A6} (\bibinfo {year} {2020})},\ \bibinfo {note} {[Erratum: Astron.Astrophys. 652, C4 (2021)]},\ \Eprint {https://arxiv.org/abs/1807.06209} {arXiv:1807.06209 [astro-ph.CO]} \BibitemShut {NoStop}%
  \bibitem [{\citenamefont {Riess}\ \emph {et~al.}(2022)\citenamefont {Riess} \emph {et~al.}}]{Riess:2021jrx}%
    \BibitemOpen
    \bibfield  {author} {\bibinfo {author} {\bibfnamefont {A.~G.}\ \bibnamefont {Riess}} \emph {et~al.},\ }\bibfield  {title} {\bibinfo {title} {{A Comprehensive Measurement of the Local Value of the Hubble Constant with 1 km s$^{-1}$ Mpc$^{-1}$ Uncertainty from the Hubble Space Telescope and the SH0ES Team}},\ }\href {https://doi.org/10.3847/2041-8213/ac5c5b} {\bibfield  {journal} {\bibinfo  {journal} {Astrophys. J. Lett.}\ }\textbf {\bibinfo {volume} {934}},\ \bibinfo {pages} {L7} (\bibinfo {year} {2022})},\ \Eprint {https://arxiv.org/abs/2112.04510} {arXiv:2112.04510 [astro-ph.CO]} \BibitemShut {NoStop}%
  \bibitem [{\citenamefont {Soares-Santos}\ \emph {et~al.}(2019)\citenamefont {Soares-Santos} \emph {et~al.}}]{DES:2019ccw}%
    \BibitemOpen
    \bibfield  {author} {\bibinfo {author} {\bibfnamefont {M.}~\bibnamefont {Soares-Santos}} \emph {et~al.} (\bibinfo {collaboration} {DES, LIGO Scientific, Virgo}),\ }\bibfield  {title} {\bibinfo {title} {{First Measurement of the Hubble Constant from a Dark Standard Siren using the Dark Energy Survey Galaxies and the LIGO/Virgo Binary\textendash{}Black-hole Merger GW170814}},\ }\href {https://doi.org/10.3847/2041-8213/ab14f1} {\bibfield  {journal} {\bibinfo  {journal} {Astrophys. J. Lett.}\ }\textbf {\bibinfo {volume} {876}},\ \bibinfo {pages} {L7} (\bibinfo {year} {2019})},\ \Eprint {https://arxiv.org/abs/1901.01540} {arXiv:1901.01540 [astro-ph.CO]} \BibitemShut {NoStop}%
  \bibitem [{\citenamefont {Palmese}\ \emph {et~al.}(2020)\citenamefont {Palmese} \emph {et~al.}}]{DES:2020nay}%
    \BibitemOpen
    \bibfield  {author} {\bibinfo {author} {\bibfnamefont {A.}~\bibnamefont {Palmese}} \emph {et~al.} (\bibinfo {collaboration} {DES}),\ }\bibfield  {title} {\bibinfo {title} {{A statistical standard siren measurement of the Hubble constant from the LIGO/Virgo gravitational wave compact object merger GW190814 and Dark Energy Survey galaxies}},\ }\href {https://doi.org/10.3847/2041-8213/abaeff} {\bibfield  {journal} {\bibinfo  {journal} {Astrophys. J. Lett.}\ }\textbf {\bibinfo {volume} {900}},\ \bibinfo {pages} {L33} (\bibinfo {year} {2020})},\ \Eprint {https://arxiv.org/abs/2006.14961} {arXiv:2006.14961 [astro-ph.CO]} \BibitemShut {NoStop}%
  \bibitem [{\citenamefont {Abbott}\ \emph {et~al.}(2021{\natexlab{d}})\citenamefont {Abbott} \emph {et~al.}}]{LIGOScientific:2019zcs}%
    \BibitemOpen
    \bibfield  {author} {\bibinfo {author} {\bibfnamefont {B.~P.}\ \bibnamefont {Abbott}} \emph {et~al.} (\bibinfo {collaboration} {LIGO Scientific, Virgo, VIRGO}),\ }\bibfield  {title} {\bibinfo {title} {{A Gravitational-wave Measurement of the Hubble Constant Following the Second Observing Run of Advanced LIGO and Virgo}},\ }\href {https://doi.org/10.3847/1538-4357/abdcb7} {\bibfield  {journal} {\bibinfo  {journal} {Astrophys. J.}\ }\textbf {\bibinfo {volume} {909}},\ \bibinfo {pages} {218} (\bibinfo {year} {2021}{\natexlab{d}})},\ \Eprint {https://arxiv.org/abs/1908.06060} {arXiv:1908.06060 [astro-ph.CO]} \BibitemShut {NoStop}%
  \bibitem [{\citenamefont {Abbott}\ \emph {et~al.}(2021{\natexlab{e}})\citenamefont {Abbott} \emph {et~al.}}]{LIGOScientific:2021aug}%
    \BibitemOpen
    \bibfield  {author} {\bibinfo {author} {\bibfnamefont {R.}~\bibnamefont {Abbott}} \emph {et~al.} (\bibinfo {collaboration} {LIGO Scientific, VIRGO, KAGRA}),\ }\bibfield  {title} {\bibinfo {title} {{Constraints on the cosmic expansion history from GWTC-3}},\ }\Eprint {https://arxiv.org/abs/2111.03604} {arXiv:2111.03604 [astro-ph.CO]}  (\bibinfo {year} {2021}{\natexlab{e}})\BibitemShut {NoStop}%
  \bibitem [{\citenamefont {Finke}\ \emph {et~al.}(2021)\citenamefont {Finke}, \citenamefont {Foffa}, \citenamefont {Iacovelli}, \citenamefont {Maggiore},\ and\ \citenamefont {Mancarella}}]{Finke:2021aom}%
    \BibitemOpen
    \bibfield  {author} {\bibinfo {author} {\bibfnamefont {A.}~\bibnamefont {Finke}}, \bibinfo {author} {\bibfnamefont {S.}~\bibnamefont {Foffa}}, \bibinfo {author} {\bibfnamefont {F.}~\bibnamefont {Iacovelli}}, \bibinfo {author} {\bibfnamefont {M.}~\bibnamefont {Maggiore}},\ and\ \bibinfo {author} {\bibfnamefont {M.}~\bibnamefont {Mancarella}},\ }\bibfield  {title} {\bibinfo {title} {{Cosmology with LIGO/Virgo dark sirens: Hubble parameter and modified gravitational wave propagation}},\ }\href {https://doi.org/10.1088/1475-7516/2021/08/026} {\bibfield  {journal} {\bibinfo  {journal} {Journal of Cosmology and Astroparticle Physics}\ }\textbf {\bibinfo {volume} {08}},\ \bibinfo {pages} {026}},\ \Eprint {https://arxiv.org/abs/2101.12660} {arXiv:2101.12660 [astro-ph.CO]} \BibitemShut {NoStop}%
  \bibitem [{\citenamefont {Yang}\ \emph {et~al.}(2022)\citenamefont {Yang}, \citenamefont {Cai}, \citenamefont {Cao},\ and\ \citenamefont {Lee}}]{Yang:2022tig}%
    \BibitemOpen
    \bibfield  {author} {\bibinfo {author} {\bibfnamefont {T.}~\bibnamefont {Yang}}, \bibinfo {author} {\bibfnamefont {R.-G.}\ \bibnamefont {Cai}}, \bibinfo {author} {\bibfnamefont {Z.}~\bibnamefont {Cao}},\ and\ \bibinfo {author} {\bibfnamefont {H.~M.}\ \bibnamefont {Lee}},\ }\bibfield  {title} {\bibinfo {title} {{Eccentricity of Long Inspiraling Compact Binaries Sheds Light on Dark Sirens}},\ }\href {https://doi.org/10.1103/PhysRevLett.129.191102} {\bibfield  {journal} {\bibinfo  {journal} {Phys. Rev. Lett.}\ }\textbf {\bibinfo {volume} {129}},\ \bibinfo {pages} {191102} (\bibinfo {year} {2022})},\ \Eprint {https://arxiv.org/abs/2202.08608} {arXiv:2202.08608 [gr-qc]} \BibitemShut {NoStop}%
  \bibitem [{\citenamefont {Liu}\ \emph {et~al.}(2020)\citenamefont {Liu}, \citenamefont {Shao}, \citenamefont {Zhao},\ and\ \citenamefont {Gao}}]{Liu:2020nwz}%
    \BibitemOpen
    \bibfield  {author} {\bibinfo {author} {\bibfnamefont {C.}~\bibnamefont {Liu}}, \bibinfo {author} {\bibfnamefont {L.}~\bibnamefont {Shao}}, \bibinfo {author} {\bibfnamefont {J.}~\bibnamefont {Zhao}},\ and\ \bibinfo {author} {\bibfnamefont {Y.}~\bibnamefont {Gao}},\ }\bibfield  {title} {\bibinfo {title} {{Multiband Observation of LIGO/Virgo Binary Black Hole Mergers in the Gravitational-wave Transient Catalog GWTC-1}},\ }\href {https://doi.org/10.1093/mnras/staa1512} {\bibfield  {journal} {\bibinfo  {journal} {Mon. Not. Roy. Astron. Soc.}\ }\textbf {\bibinfo {volume} {496}},\ \bibinfo {pages} {182} (\bibinfo {year} {2020})},\ \Eprint {https://arxiv.org/abs/2004.12096} {arXiv:2004.12096 [astro-ph.HE]} \BibitemShut {NoStop}%
  \bibitem [{\citenamefont {Buonanno}\ \emph {et~al.}(2009)\citenamefont {Buonanno}, \citenamefont {Iyer}, \citenamefont {Ochsner}, \citenamefont {Pan},\ and\ \citenamefont {Sathyaprakash}}]{Buonanno:2009zt}%
    \BibitemOpen
    \bibfield  {author} {\bibinfo {author} {\bibfnamefont {A.}~\bibnamefont {Buonanno}}, \bibinfo {author} {\bibfnamefont {B.}~\bibnamefont {Iyer}}, \bibinfo {author} {\bibfnamefont {E.}~\bibnamefont {Ochsner}}, \bibinfo {author} {\bibfnamefont {Y.}~\bibnamefont {Pan}},\ and\ \bibinfo {author} {\bibfnamefont {B.~S.}\ \bibnamefont {Sathyaprakash}},\ }\bibfield  {title} {\bibinfo {title} {{Comparison of post-Newtonian templates for compact binary inspiral signals in gravitational-wave detectors}},\ }\href {https://doi.org/10.1103/PhysRevD.80.084043} {\bibfield  {journal} {\bibinfo  {journal} {Phys. Rev. D}\ }\textbf {\bibinfo {volume} {80}},\ \bibinfo {pages} {084043} (\bibinfo {year} {2009})},\ \Eprint {https://arxiv.org/abs/0907.0700} {arXiv:0907.0700 [gr-qc]} \BibitemShut {NoStop}%
  \bibitem [{\citenamefont {Buonanno}\ \emph {et~al.}(2007)\citenamefont {Buonanno}, \citenamefont {Cook},\ and\ \citenamefont {Pretorius}}]{Buonanno:2006ui}%
    \BibitemOpen
    \bibfield  {author} {\bibinfo {author} {\bibfnamefont {A.}~\bibnamefont {Buonanno}}, \bibinfo {author} {\bibfnamefont {G.~B.}\ \bibnamefont {Cook}},\ and\ \bibinfo {author} {\bibfnamefont {F.}~\bibnamefont {Pretorius}},\ }\bibfield  {title} {\bibinfo {title} {{Inspiral, merger and ring-down of equal-mass black-hole binaries}},\ }\href {https://doi.org/10.1103/PhysRevD.75.124018} {\bibfield  {journal} {\bibinfo  {journal} {Phys. Rev. D}\ }\textbf {\bibinfo {volume} {75}},\ \bibinfo {pages} {124018} (\bibinfo {year} {2007})},\ \Eprint {https://arxiv.org/abs/gr-qc/0610122} {arXiv:gr-qc/0610122} \BibitemShut {NoStop}%
  \bibitem [{\citenamefont {Finn}(1992)}]{Finn:1992wt}%
    \BibitemOpen
    \bibfield  {author} {\bibinfo {author} {\bibfnamefont {L.~S.}\ \bibnamefont {Finn}},\ }\bibfield  {title} {\bibinfo {title} {{Detection, measurement and gravitational radiation}},\ }\href {https://doi.org/10.1103/PhysRevD.46.5236} {\bibfield  {journal} {\bibinfo  {journal} {Phys. Rev. D}\ }\textbf {\bibinfo {volume} {46}},\ \bibinfo {pages} {5236} (\bibinfo {year} {1992})},\ \Eprint {https://arxiv.org/abs/gr-qc/9209010} {arXiv:gr-qc/9209010} \BibitemShut {NoStop}%
  \bibitem [{\citenamefont {Robson}\ \emph {et~al.}(2019)\citenamefont {Robson}, \citenamefont {Cornish},\ and\ \citenamefont {Liu}}]{Robson:2018ifk}%
    \BibitemOpen
    \bibfield  {author} {\bibinfo {author} {\bibfnamefont {T.}~\bibnamefont {Robson}}, \bibinfo {author} {\bibfnamefont {N.~J.}\ \bibnamefont {Cornish}},\ and\ \bibinfo {author} {\bibfnamefont {C.}~\bibnamefont {Liu}},\ }\bibfield  {title} {\bibinfo {title} {{The construction and use of LISA sensitivity curves}},\ }\href {https://doi.org/10.1088/1361-6382/ab1101} {\bibfield  {journal} {\bibinfo  {journal} {Class. Quant. Grav.}\ }\textbf {\bibinfo {volume} {36}},\ \bibinfo {pages} {105011} (\bibinfo {year} {2019})},\ \Eprint {https://arxiv.org/abs/1803.01944} {arXiv:1803.01944 [astro-ph.HE]} \BibitemShut {NoStop}%
  \bibitem [{\citenamefont {Mills}\ and\ \citenamefont {Fairhurst}(2021)}]{Mills:2020thr}%
    \BibitemOpen
    \bibfield  {author} {\bibinfo {author} {\bibfnamefont {C.}~\bibnamefont {Mills}}\ and\ \bibinfo {author} {\bibfnamefont {S.}~\bibnamefont {Fairhurst}},\ }\bibfield  {title} {\bibinfo {title} {{Measuring gravitational-wave higher-order multipoles}},\ }\href {https://doi.org/10.1103/PhysRevD.103.024042} {\bibfield  {journal} {\bibinfo  {journal} {Phys. Rev. D}\ }\textbf {\bibinfo {volume} {103}},\ \bibinfo {pages} {024042} (\bibinfo {year} {2021})},\ \Eprint {https://arxiv.org/abs/2007.04313} {arXiv:2007.04313 [gr-qc]} \BibitemShut {NoStop}%
  \bibitem [{\citenamefont {Veitch}\ \emph {et~al.}(2015)\citenamefont {Veitch} \emph {et~al.}}]{Veitch:2014wba}%
    \BibitemOpen
    \bibfield  {author} {\bibinfo {author} {\bibfnamefont {J.}~\bibnamefont {Veitch}} \emph {et~al.},\ }\bibfield  {title} {\bibinfo {title} {{Parameter estimation for compact binaries with ground-based gravitational-wave observations using the LALInference software library}},\ }\href {https://doi.org/10.1103/PhysRevD.91.042003} {\bibfield  {journal} {\bibinfo  {journal} {Phys. Rev. D}\ }\textbf {\bibinfo {volume} {91}},\ \bibinfo {pages} {042003} (\bibinfo {year} {2015})},\ \Eprint {https://arxiv.org/abs/1409.7215} {arXiv:1409.7215 [gr-qc]} \BibitemShut {NoStop}%
  \bibitem [{\citenamefont {Vallisneri}(2008)}]{Vallisneri:2007ev}%
    \BibitemOpen
    \bibfield  {author} {\bibinfo {author} {\bibfnamefont {M.}~\bibnamefont {Vallisneri}},\ }\bibfield  {title} {\bibinfo {title} {{Use and abuse of the Fisher information matrix in the assessment of gravitational-wave parameter-estimation prospects}},\ }\href {https://doi.org/10.1103/PhysRevD.77.042001} {\bibfield  {journal} {\bibinfo  {journal} {Phys. Rev. D}\ }\textbf {\bibinfo {volume} {77}},\ \bibinfo {pages} {042001} (\bibinfo {year} {2008})},\ \Eprint {https://arxiv.org/abs/gr-qc/0703086} {arXiv:gr-qc/0703086} \BibitemShut {NoStop}%
  \bibitem [{\citenamefont {Zhao}\ \emph {et~al.}(2021)\citenamefont {Zhao}, \citenamefont {Shao}, \citenamefont {Gao}, \citenamefont {Liu}, \citenamefont {Cao},\ and\ \citenamefont {Ma}}]{Zhao:2021bjw}%
    \BibitemOpen
    \bibfield  {author} {\bibinfo {author} {\bibfnamefont {J.}~\bibnamefont {Zhao}}, \bibinfo {author} {\bibfnamefont {L.}~\bibnamefont {Shao}}, \bibinfo {author} {\bibfnamefont {Y.}~\bibnamefont {Gao}}, \bibinfo {author} {\bibfnamefont {C.}~\bibnamefont {Liu}}, \bibinfo {author} {\bibfnamefont {Z.}~\bibnamefont {Cao}},\ and\ \bibinfo {author} {\bibfnamefont {B.-Q.}\ \bibnamefont {Ma}},\ }\bibfield  {title} {\bibinfo {title} {{Probing dipole radiation from binary neutron stars with ground-based laser-interferometer and atom-interferometer gravitational-wave observatories}},\ }\href {https://doi.org/10.1103/PhysRevD.104.084008} {\bibfield  {journal} {\bibinfo  {journal} {Phys. Rev. D}\ }\textbf {\bibinfo {volume} {104}},\ \bibinfo {pages} {084008} (\bibinfo {year} {2021})},\ \Eprint {https://arxiv.org/abs/2106.04883} {arXiv:2106.04883 [gr-qc]} \BibitemShut {NoStop}%
  \bibitem [{\citenamefont {Cutler}\ and\ \citenamefont {Flanagan}(1994)}]{Cutler:1994ys}%
    \BibitemOpen
    \bibfield  {author} {\bibinfo {author} {\bibfnamefont {C.}~\bibnamefont {Cutler}}\ and\ \bibinfo {author} {\bibfnamefont {E.~E.}\ \bibnamefont {Flanagan}},\ }\bibfield  {title} {\bibinfo {title} {{Gravitational waves from merging compact binaries: How accurately can one extract the binary's parameters from the inspiral wave form?}},\ }\href {https://doi.org/10.1103/PhysRevD.49.2658} {\bibfield  {journal} {\bibinfo  {journal} {Phys. Rev. D}\ }\textbf {\bibinfo {volume} {49}},\ \bibinfo {pages} {2658} (\bibinfo {year} {1994})},\ \Eprint {https://arxiv.org/abs/gr-qc/9402014} {arXiv:gr-qc/9402014} \BibitemShut {NoStop}%
  \bibitem [{\citenamefont {Bradbury}\ \emph {et~al.}(2018)\citenamefont {Bradbury}, \citenamefont {Frostig}, \citenamefont {Hawkins}, \citenamefont {Johnson}, \citenamefont {Leary}, \citenamefont {Maclaurin}, \citenamefont {Necula}, \citenamefont {Paszke}, \citenamefont {Vander{P}las}, \citenamefont {Wanderman-{M}ilne},\ and\ \citenamefont {Zhang}}]{JAX:2018Github}%
    \BibitemOpen
    \bibfield  {author} {\bibinfo {author} {\bibfnamefont {J.}~\bibnamefont {Bradbury}}, \bibinfo {author} {\bibfnamefont {R.}~\bibnamefont {Frostig}}, \bibinfo {author} {\bibfnamefont {P.}~\bibnamefont {Hawkins}}, \bibinfo {author} {\bibfnamefont {M.~J.}\ \bibnamefont {Johnson}}, \bibinfo {author} {\bibfnamefont {C.}~\bibnamefont {Leary}}, \bibinfo {author} {\bibfnamefont {D.}~\bibnamefont {Maclaurin}}, \bibinfo {author} {\bibfnamefont {G.}~\bibnamefont {Necula}}, \bibinfo {author} {\bibfnamefont {A.}~\bibnamefont {Paszke}}, \bibinfo {author} {\bibfnamefont {J.}~\bibnamefont {Vander{P}las}}, \bibinfo {author} {\bibfnamefont {S.}~\bibnamefont {Wanderman-{M}ilne}},\ and\ \bibinfo {author} {\bibfnamefont {Q.}~\bibnamefont {Zhang}},\ }\href {http://github.com/google/jax} {\bibinfo {title} {{JAX}: composable transformations of {P}ython+{N}um{P}y programs}} (\bibinfo {year} {2018})\BibitemShut {NoStop}%
  \bibitem [{\citenamefont {Barack}\ and\ \citenamefont {Cutler}(2004)}]{Barack:2003fp}%
    \BibitemOpen
    \bibfield  {author} {\bibinfo {author} {\bibfnamefont {L.}~\bibnamefont {Barack}}\ and\ \bibinfo {author} {\bibfnamefont {C.}~\bibnamefont {Cutler}},\ }\bibfield  {title} {\bibinfo {title} {{LISA capture sources: Approximate waveforms, signal-to-noise ratios, and parameter estimation accuracy}},\ }\href {https://doi.org/10.1103/PhysRevD.69.082005} {\bibfield  {journal} {\bibinfo  {journal} {Phys. Rev. D}\ }\textbf {\bibinfo {volume} {69}},\ \bibinfo {pages} {082005} (\bibinfo {year} {2004})},\ \Eprint {https://arxiv.org/abs/gr-qc/0310125} {arXiv:gr-qc/0310125} \BibitemShut {NoStop}%
  \bibitem [{\citenamefont {Kopparapu}\ \emph {et~al.}(2008)\citenamefont {Kopparapu}, \citenamefont {Hanna}, \citenamefont {Kalogera}, \citenamefont {O'Shaughnessy}, \citenamefont {Gonz\'alez}, \citenamefont {Brady},\ and\ \citenamefont {Fairhurst}}]{Kopparapu:2007ib}%
    \BibitemOpen
    \bibfield  {author} {\bibinfo {author} {\bibfnamefont {R.~K.}\ \bibnamefont {Kopparapu}}, \bibinfo {author} {\bibfnamefont {C.}~\bibnamefont {Hanna}}, \bibinfo {author} {\bibfnamefont {V.}~\bibnamefont {Kalogera}}, \bibinfo {author} {\bibfnamefont {R.}~\bibnamefont {O'Shaughnessy}}, \bibinfo {author} {\bibfnamefont {G.}~\bibnamefont {Gonz\'alez}}, \bibinfo {author} {\bibfnamefont {P.~R.}\ \bibnamefont {Brady}},\ and\ \bibinfo {author} {\bibfnamefont {S.}~\bibnamefont {Fairhurst}},\ }\bibfield  {title} {\bibinfo {title} {{Host Galaxies Catalog Used in LIGO Searches for Compact Binary Coalescence Events}},\ }\href {https://doi.org/10.1086/527348} {\bibfield  {journal} {\bibinfo  {journal} {Astrophys. J.}\ }\textbf {\bibinfo {volume} {675}},\ \bibinfo {pages} {1459} (\bibinfo {year} {2008})},\ \Eprint {https://arxiv.org/abs/0706.1283} {arXiv:0706.1283 [astro-ph]} \BibitemShut {NoStop}%
  \bibitem [{\citenamefont {Abadie}\ \emph {et~al.}(2010)\citenamefont {Abadie} \emph {et~al.}}]{LIGOScientific:2010nhs}%
    \BibitemOpen
    \bibfield  {author} {\bibinfo {author} {\bibfnamefont {J.}~\bibnamefont {Abadie}} \emph {et~al.} (\bibinfo {collaboration} {LIGO Scientific, VIRGO}),\ }\bibfield  {title} {\bibinfo {title} {{Predictions for the Rates of Compact Binary Coalescences Observable by Ground-based Gravitational-wave Detectors}},\ }\href {https://doi.org/10.1088/0264-9381/27/17/173001} {\bibfield  {journal} {\bibinfo  {journal} {Class. Quant. Grav.}\ }\textbf {\bibinfo {volume} {27}},\ \bibinfo {pages} {173001} (\bibinfo {year} {2010})},\ \Eprint {https://arxiv.org/abs/1003.2480} {arXiv:1003.2480 [astro-ph.HE]} \BibitemShut {NoStop}%
  \bibitem [{\citenamefont {Chen}\ and\ \citenamefont {Holz}(2016)}]{Chen:2016tys}%
    \BibitemOpen
    \bibfield  {author} {\bibinfo {author} {\bibfnamefont {H.-Y.}\ \bibnamefont {Chen}}\ and\ \bibinfo {author} {\bibfnamefont {D.~E.}\ \bibnamefont {Holz}},\ }\bibfield  {title} {\bibinfo {title} {{Finding the One: Identifying the Host Galaxies of Gravitational-Wave Sources}},\ }\Eprint {https://arxiv.org/abs/1612.01471} {arXiv:1612.01471 [astro-ph.HE]}  (\bibinfo {year} {2016})\BibitemShut {NoStop}%
  \bibitem [{\citenamefont {Bulla}\ \emph {et~al.}(2022)\citenamefont {Bulla}, \citenamefont {Coughlin}, \citenamefont {Dhawan},\ and\ \citenamefont {Dietrich}}]{Bulla:2022ppy}%
    \BibitemOpen
    \bibfield  {author} {\bibinfo {author} {\bibfnamefont {M.}~\bibnamefont {Bulla}}, \bibinfo {author} {\bibfnamefont {M.~W.}\ \bibnamefont {Coughlin}}, \bibinfo {author} {\bibfnamefont {S.}~\bibnamefont {Dhawan}},\ and\ \bibinfo {author} {\bibfnamefont {T.}~\bibnamefont {Dietrich}},\ }\bibfield  {title} {\bibinfo {title} {{Multi-Messenger Constraints on the Hubble Constant through Combination of Gravitational Waves, Gamma-Ray Bursts and Kilonovae from Neutron Star Mergers}},\ }\href {https://doi.org/10.3390/universe8050289} {\bibfield  {journal} {\bibinfo  {journal} {Universe}\ }\textbf {\bibinfo {volume} {8}},\ \bibinfo {pages} {289} (\bibinfo {year} {2022})},\ \Eprint {https://arxiv.org/abs/2205.09145} {arXiv:2205.09145 [astro-ph.HE]} \BibitemShut {NoStop}%
  \bibitem [{\citenamefont {Wu}\ \emph {et~al.}(2021)\citenamefont {Wu} \emph {et~al.}}]{TaijiScientific:2021qgx}%
    \BibitemOpen
    \bibfield  {author} {\bibinfo {author} {\bibfnamefont {Y.-L.}\ \bibnamefont {Wu}} \emph {et~al.} (\bibinfo {collaboration} {Taiji Scientific}),\ }\bibfield  {title} {\bibinfo {title} {{China\textquoteright{}s first step towards probing the expanding universe and the nature of gravity using a space borne gravitational wave antenna}},\ }\href {https://doi.org/10.1038/s42005-021-00529-z} {\bibfield  {journal} {\bibinfo  {journal} {Commun. Phys.}\ }\textbf {\bibinfo {volume} {4}},\ \bibinfo {pages} {34} (\bibinfo {year} {2021})}\BibitemShut {NoStop}%
  \bibitem [{\citenamefont {Luo}\ \emph {et~al.}(2016)\citenamefont {Luo} \emph {et~al.}}]{TianQin:2015yph}%
    \BibitemOpen
    \bibfield  {author} {\bibinfo {author} {\bibfnamefont {J.}~\bibnamefont {Luo}} \emph {et~al.} (\bibinfo {collaboration} {TianQin}),\ }\bibfield  {title} {\bibinfo {title} {{TianQin: a space-borne gravitational wave detector}},\ }\href {https://doi.org/10.1088/0264-9381/33/3/035010} {\bibfield  {journal} {\bibinfo  {journal} {Class. Quant. Grav.}\ }\textbf {\bibinfo {volume} {33}},\ \bibinfo {pages} {035010} (\bibinfo {year} {2016})},\ \Eprint {https://arxiv.org/abs/1512.02076} {arXiv:1512.02076 [astro-ph.IM]} \BibitemShut {NoStop}%
  \bibitem [{\citenamefont {Sato}\ \emph {et~al.}(2017)\citenamefont {Sato} \emph {et~al.}}]{Sato:2017dkf}%
    \BibitemOpen
    \bibfield  {author} {\bibinfo {author} {\bibfnamefont {S.}~\bibnamefont {Sato}} \emph {et~al.},\ }\bibfield  {title} {\bibinfo {title} {{The status of DECIGO}},\ }\href {https://doi.org/10.1088/1742-6596/840/1/012010} {\bibfield  {journal} {\bibinfo  {journal} {J. Phys. Conf. Ser.}\ }\textbf {\bibinfo {volume} {840}},\ \bibinfo {pages} {012010} (\bibinfo {year} {2017})}\BibitemShut {NoStop}%
  \end{thebibliography}
%
  

\end{document}